\documentclass[12pt]{aastex62}

\begin{document}

\parindent=1.0cm

\title{V1507 CYGNI (HD187399): A HIGHLY EVOLVED, ENIGMATIC INTERACTING BINARY SYSTEM WITH AN ECCENTRIC ORBIT}

\author{T. J. Davidge}

\affiliation{Dominion Astrophysical Observatory,
\\Herzberg Astronomy \& Astrophysics Research Center,
\\National Research Council of Canada, 5071 West Saanich Road,
\\Victoria, BC Canada V9E 2E7\\tim.davidge@nrc.ca; tdavidge1450@gmail.com}

\begin{abstract}

	The properties of the interacting, eccentric orbit binary V1507 Cyg 
(HD187399) are examined with spectra that cover wavelengths from 0.63 to 
$0.68\mu$m. The spectrum of the brightest star is very similar to that of 
the B8 I star $\beta$ Ori, although with absorption lines that show 
sub-structure consistent with a varying tidal field. The bulk of the 
H$\alpha$ emission in the spectrum appears to be associated 
with the brighter star. Evidence is presented that 
the period of the system has been stable over timescales of many 
decades, arguing against large-scale mass transfer at the current epoch. 
Absorption and emission lines are identified that originate in an expanding 
asymmetric envelope around the companion, and 
component masses of $6.4 \pm 0.9$ and $14.0 \pm 0.9$ M$_{\odot}$ are found, 
where the former applies to the brighter star and an 
inclination of 46$^o$ has been assumed. The evolutionary state of V1507 Cyg 
is then that of a system in which mass transfer has progressed to the 
point where the mass ratio has been reversed, with the 
brighter star stripped of much of its initial mass. 
It is argued that the brighter star is an $\alpha$ Cyg 
variable, and that it is those light variations that dominate the 
system light curve. V1507 Cyg is observed at or near the center of a diffuse HI 
bubble that is detected at 408 and 1420MHz. It is suggested that the eccentric 
orbit is the result of evolution in a hierarchical system, in which 
a now-defunct massive third body recently disturbed the orbit of the 
stars in V1507 Cyg, thereby disrupting mass transfer.

\end{abstract}

\section{INTRODUCTION}

	Binary stars play an important role in astrophysics. Non-interacting 
systems, in which the components are well detached, serve as fundamental 
calibrators of basic stellar properties \citep[e.g.][]{toretal2010}, and are 
also primary distance indicators \citep[e.g.][]{dev1978, renetal2021, sta2021}. 
As for interacting systems, these are relatively common in the 
solar neighborhood, and mass transfer is thought to produce well-studied 
objects such as short period contact binaries \citep[e.g.][]{ste2006} 
and semi-detached systems \citep[]{mor1960, sma1962}. The former are also 
distance indicators \citep[e.g.][]{ruc1997}. Interactions 
between very massive stars may also produce some of the most luminous objects 
in a galaxy, such as luminous blue variables (LBVs) and B[e] stars 
\citep[e.g.][]{claetal2013, smi2015}. If the initial conditions are right then 
the end result of interacting binary evolution will be the merger of stellar 
remnants within a Hubble time, resulting in the production of elements such as 
Au \citep[e.g.][]{kobetal2020}. The immediate progenitors of such 
exotic systems have been detected in the solar neighborhood 
\citep[e.g.][]{buretal2019, buretal2020}. 

	The final state of a binary system is influenced by many factors, and 
so there is an obvious interest in characterizing the properties of a diverse 
population of close binary systems (CBSs), with a goal of 
linking the formation and early evolution of various system types with 
final end states. Spectra of the CBS V1507 Cyg (HD187399) are 
discussed in the present paper. V1507 Cyg was observed as part of a 
program underway with the 1.2 metre telescope at the Dominion Astrophysical 
Observatory (DAO) to record spectra of binaries with periods of tens of days 
that are thought to be in the early stages of mass transfer. Such systems are 
of interest since if mass transfer is not well-advanced then the component 
stars and system properties may not yet have departed far from their 
initial states, simplifying the task of characterizing the progenitors. This 
is also an evolutionary stage where the mass flow rate is expected to be 
very high, and significant amounts of mass may be lost from the system. 

	V1507 Cyg is in a list of possible W Ser systems and related objects 
that was compiled by \citet{wiletal1984}. They state that V1507 Cyg is 
{\it 'possibly related to W Ser stars, possibly not. Strange case.'} 
The W Ser stars were first recognized as a group by 
\citet{pla1978}, who identified six CBSs that have unusually high 
levels of UV emission. This emission, as well as other observational properties 
of W Ser systems, is thought to be the result of large scale 
mass transfer from the originally more massive star to a disk around 
the companion.

	The binary nature of V1507 Cyg was first discussed by \citet{mer1949}, 
who examined spectra that span a multi-decade time line to show 
that it is a single line spectroscopic binary at visible wavelengths with 
a moderately eccentric orbit. \citet{mer1949} concluded that the spectrum 
contains contributions from stellar and nonstellar sources. 
Signs of activity within the system were also identified. For example, 
lines of Ca were found to have strengths and velocities that vary 
with time in a complex manner. 

	Interest in V1507 Cyg was piqued when 
\citet{tri1969} suggested that it might host a collapsed object 
based on the large mass function found by \citet{mer1949}. An appealing aspect 
of this suggestion is that a collapsed object could explain the 
eccentric orbit of the system, as the supernova that would have produced 
it could alter the orbits of the component stars \citep[e.g.][]{maretal2009}. 
Motivated by the potential of finding a collapsed 
object, \citet{hut1972} and \citet{hut1973} 
examined the V1507 Cyg spectrum, and found signs of 
activity in the system. The Balmer lines were found to have a P Cygni profile, 
indicating an outward flow of material. Possible variability 
in the properties of H$\alpha$ on time scales of 
minutes \citep[]{hut1973} was also detected, suggesting  that there 
are major changes in the intrasystem environment over modest spatial scales. 
\citet{hut1973} conclude that the unseen companion is likely not 
a collapsed object given the absence of X-ray emission. 

	\citet{hil1976} and \citet{pavetal1979} find that 
the light curve of V1507 Cyg follows a wave-like shape 
with no evidence of eclipses, indicating an 
orbital inclination that is less than $\sim 60^o$. 
The light curve shape is stable on the time scale of many 
years, although there is a cycle-to-cycle dispersion in the measurements. 
Later in this paper it is argued that the light curve is dominated by 
the intrinsic variability of the brighter star in the system.

	\citet{ber1998} discuss polarimetric 
observations of V1507 Cyg. An orbital inclination of $46 \pm 
2$ degrees was found after de-coupling the interstellar 
and system contributions. This orientation is consistent 
with the absence of eclipses and the lack of complex large-scale 
photometric variations if light from accretion activity 
is largely restricted to the orbital plane. \citet{ber1998} also 
note that a collapsed object is not required to explain the system properties.

	\citet{dav2023} used a sample of stars selected from the GAIA 
DR3 \citep[]{gai2022} to search for signatures of a fossil cluster 
or moving group that might be associated with V1507 Cyg. The parallax of 
V1507 Cyg indicates that it does not belong to the more distant young 
clusters and star-forming regions that are seen close to it on the 
sky. Still, there are hints that it might be associated with 
a diffuse cluster, in which it would be an outlier both 
in terms of physical location and proper motions. 
Based on the brightness of possible companions 
selected using a number of criteria, Davidge assigned a lower 
limit to the initial mass for the donor star of $1.8\pm 0.3$M$_{\odot}$, 
and an approximate system age of no more than 1 Gyr. A faint 
extended circumsystem envelope at mid-infrared (MIR) wavelengths 
was also found, as might be expected if there has been non-conservative 
mass transfer \citep[]{desetal2015}. 

	There are numerous outstanding questions regarding the 
nature of V1507 Cyg and its relation to other CBSs. For example, 
is there evidence for mass flow within the system? If so, how does the mass 
flow vary throughout the orbital cycle? If there has been recent mass transfer 
then there may be evidence for an accretion disk/envelope around the gaining 
star, while an eccentric orbit should cause any mass flow to be episodic. A 
more basic question is 'What are the masses of the component stars'? To date, 
V1507 Cyg has been examined in the context of a single line spectroscopic 
binary. If the motion of the companion can be traced then the masses 
of both stars can be estimated, thereby providing insights into the 
nature of the system. Finally, what is the origin of the 
eccentric orbit? An eccentric orbit is surprising for 
a CBS given that tidal effects are expected to 
rapidly circularize the orbit \citep[]{zah1977, moretal2010}, 
although there may be exceptions \citep[e.g.]{sepetal2009}. 

	In the current paper we discuss spectra of V1507 Cyg at red wavelengths 
that were recorded over multiple orbital cycles during a single observing 
season. Absorption features in the spectrum of the brighter star are 
identified, as are features from an envelope that surrounds the companion. 
The properties of the brighter star are assessed to determine if it has been 
subjected to large scale mass loss. In addition, radial velocity measurements 
of the accretion disk are used as a proxy to track the orbital motion of 
the companion star, thereby providing information about the masses of both 
components. The morphology of H$\alpha$ emission is also examined, and 
at least two distinct components are identified. Finally, maps from the 
Canadian Galactic Plane Survey \citep[CGPS][]{tayetal2003} are used to explore 
the interstellar medium (ISM) around V1507 Cyg to search for 
possible additional clues into the past history of the system.
 
\section{OBSERVATIONS}

	Spectra were recorded with the DAO 1.2 meter Petrie telescope and 
the McKellar spectrograph \citep[]{monetal2014} over the course of 36 
nights in 2022. The spectrograph was configured with 
the 32 inch camera, IS32R image slicer, the 1200H grating, and 
the $2048 \times 4088$ pixel SITE--4 CCD. 
The observations were recorded with the telescope operating in 
robotic mode, and accounted for only a modest portion of the available 
time on any given night. The central wavelength of the spectra varied by 
$\pm 100\AA$ over the course of the observations to accommodate other 
programs, and all of the V1507 Cyg spectra sample the 6300\AA\ to 6800\AA\ 
wavelength interval. In addition to H$\alpha$, this wavelength region 
contains atomic transitions of metals (mostly Si and Fe) and 
He that produce lines in the spectra of moderately early-type stars, 
accretion disks, and circumstellar/system shells. The 
wavelength resolution is $\frac{\lambda}{\Delta\lambda} \sim 17000$. 

	A complete nightly observing sequence consisted of 
three 300 sec exposures, and these were followed by 
an observation of a ThAr arc. Three 30 sec exposures 
of Vega were also recorded either before or after V1507 Cyg to monitor 
telluric absorption and sky transparency. Cloud cover was also tracked with 
an on-sky camera, and an inspection of the final spectra indicate that 
thick clouds were present during only a handful of nights. 

	The dates on which spectra were recorded and the corresponding 
orbital phases are listed in Table 1, where the latter were calculated 
using the ephemeris generated by \citet{hut1973}. 
Later in the paper it is demonstrated that this ephemeris still applies 
at the current epoch -- there is no evidence of a period change over the 
many decades that this system has now been observed. 
Nights on which clouds have a noticeable impact on the final S/N ratio of the 
processed spectra are indicated with an asterisk in Table 1.

\begin{center}
\begin{deluxetable}{llll}
\tablecaption{Dates of Observations and Phase Coverage}
\tablehead{ HJD\tablenotemark{a} & Orbital\tablenotemark{b} & HJD\tablenotemark{a} & Orbital\tablenotemark{b} \\ -- 2400000 & Phase & --2400000 & Phase }
\startdata
59684.381 & 0.111 & 59786.421 & 0.759 \\
59686.364 & 0.182 & 59818.174 & 0.894 \\
59695.414 & 0.505 & 59820.334 & 0.971 \\
59696.348 & 0.539 & 59821.310 & 0.006 \\
59698.381 & 0.611 & 59822.166 & 0.037 \\
59708.430 & 0.971 & 59824.272 & 0.105 \\
59709.366 & 0.004 & 59825.280* & 0.141 \\
59737.423 & 0.007 & 59827.256 & 0.219 \\
59751.381 & 0.506 & 59828.155 & 0.251 \\
59753.331* & 0.576 & 59829.185 & 0.288 \\
59758.344 & 0.755 & 59847.340 & 0.937 \\
59761.339 & 0.862 & 59848.252 & 0.972 \\
59762.326 & 0.897 & 59849.172 & 0.003 \\
59780.211 & 0.537 & 59850.117* & 0.036 \\
59781.211 & 0.573 & 59852.131 & 0.108 \\
59783.462* & 0.645 & 59853.109 & 0.143 \\
59784.207 & 0.680 & 59861.979 & 0.429 \\
59785.434 & 0.724 & 59862.097 & 0.465 \\
\enddata
\tablenotetext{a}{Asterisks denote nights with poor transparency.}
\tablenotetext{b}{Calculated with the \citet{hut1973} ephemeris.}
\end{deluxetable}
\end{center}

	Artifacts in the spectra introduced by the 
instrumentation and the sky were removed using 
standard procedures. These steps include (1) summing the 
three spectra recorded on each night, (2) the removal 
of the floating bias and the fixed detector bias pattern, (3) the removal 
of cosmic rays, (4) flat-fielding, (5) correcting for scattered light, and (6) 
summing the signal within the full width at half maximum (FWHM) of the 
stellar profile as defined by the image slicer. The extracted spectra were 
then wavelength calibrated and normalized to a pseudo-continuum.

\section{THE SPECTRUM OF THE BRIGHTER STAR}

	An example of a processed spectrum that has a signal 
to noise ratio that is typical of the dataset in general is shown 
in Figure 1. The dominant feature is H$\alpha$ emission, which masks any 
H$\alpha$ absorption that might originate in the spectrum of either 
star. H$\alpha$ emission can originate from a number of sources 
\citep[e.g.][]{broetal2021}, and a discussion of this feature is deferred to 
later in the paper. In this section we focus on the narrow absorption 
lines of SiII and HeI that are evident in Figure 1.

\begin{figure}
\figurenum{1}
\plotone{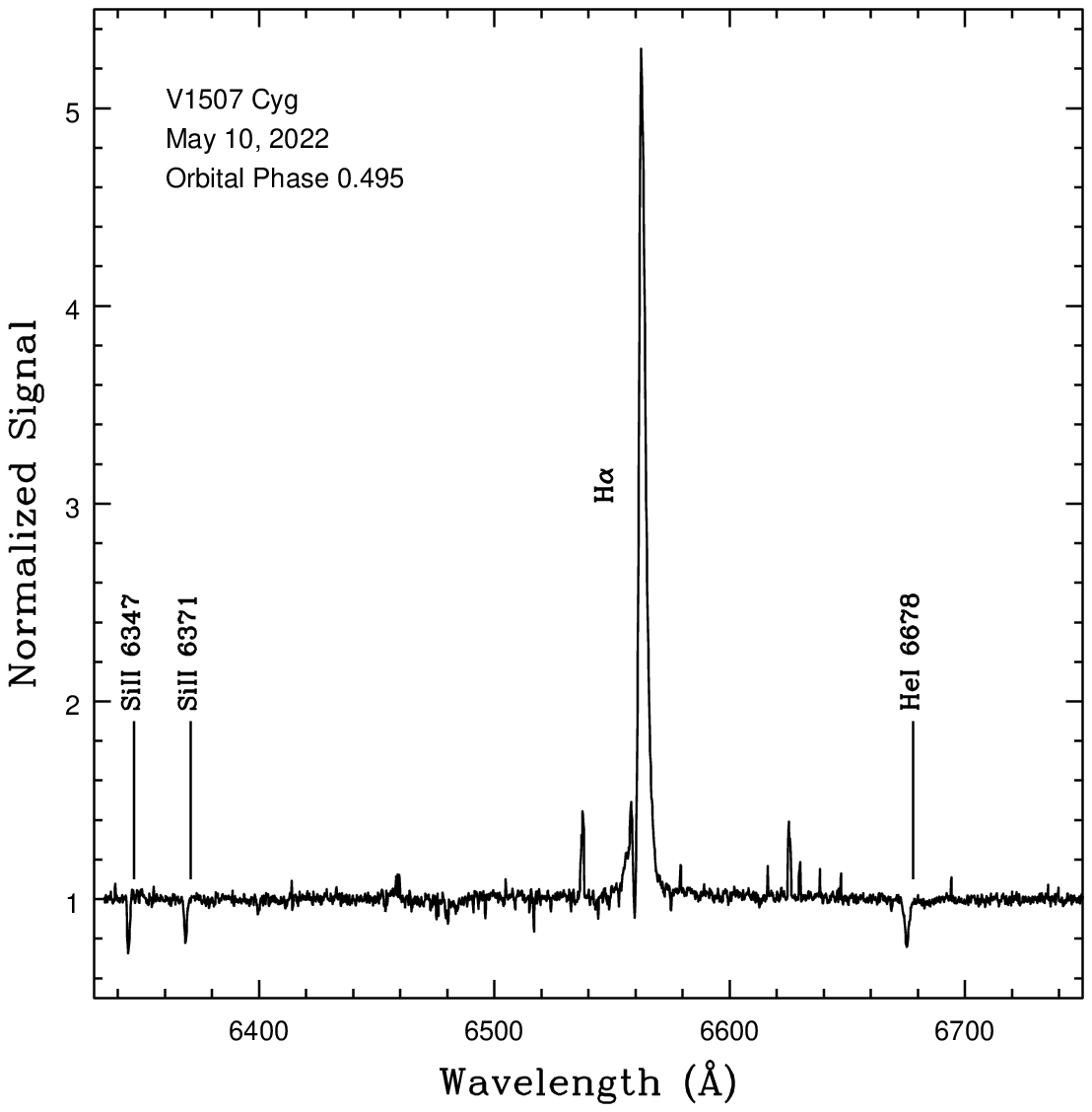}
\caption{Spectrum of V1507 Cyg that was recorded on May 10, 2022. 
The SiII and HeI lines originate mainly in the spectrum of the brighter 
star, while the H$\alpha$ emission originates in circumstellar and intrasystem 
material. The weak emission lines are cosmic-rays 
that were not suppressed during processing. The slight shift in wavelength 
between the expected and observed locations of the SiII and HeI lines is 
due to the orbital motion of the brighter star.}
\end{figure}

	Wavelength intervals that contain SiII, FeII, and HeI lines are 
shown in Figure 2. The spectra in this figure are the means of nightly 
spectra that were binned in phase intervals that cover 10\% of each 
orbital cycle. This binning is a compromise between the gain in 
signal-to-noise ratio that results from combining individual spectra, and the 
blurring of line characteristics that occurs with orbital phase. The velocity 
differential is greatest near phase 0.0, where the dispersion is 
$\pm 20$km/sec, or $\pm 0.4$\AA\ at these wavelengths.

	The wavelengths of the absorption lines vary with orbital phase, 
with the centroids of the SiII and HeI lines moving by almost 4\AA\ (i.e. 
$\sim 190$ km/sec) throughout an orbital cycle. This is in agreement with 
the range in radial velocities measured at shorter wavelengths by 
\citet{mer1949}. SiII 6347 is a resonance transition, and so is more 
susceptible to ISM absorption than SiII 6171. The properties of SiII 6347 
vary in lockstep with those of SiII 6171, indicating that the 
dominant contributor to SiII 6347 is stellar in origin.

\begin{figure}
\figurenum{2}
\plotone{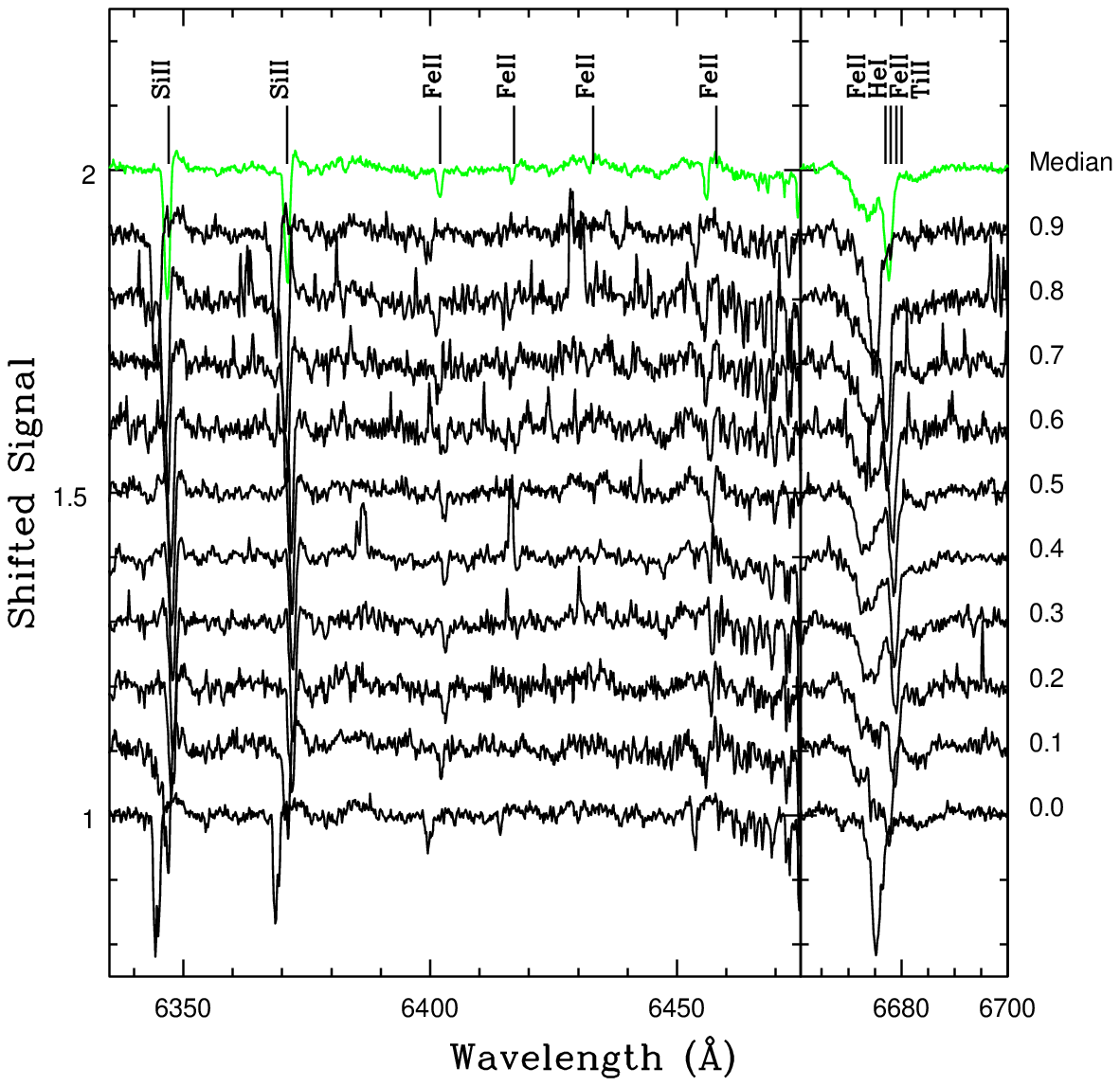}
\caption{Phase-binned spectra of V1507 Cyg in two wavelength intervals. The 
centroids of the phase bins are listed along the right axis. The spectrum 
plotted in green is the median of the binned spectra after shifting them 
into the rest frame. The shapes of the SiII and HeI lines change with orbital 
phase in a similar manner. The centers of the SiII, FeII, and HeI lines 
have a skewed appearance near phase 0.0, and in the text this behaviour is 
attributed to tidal effects during periastron. The depression in 
the spectrum at wavelengths shortward of HeI does not track the 
HeI line in the spectrum of the brighter star, and in Section 4 it is 
argued that this trough is part of a P Cygni profile 
that originates in an envelope around the heavily obscured companion.}
\end{figure}

	The strengths and shapes of the SiII and HeI lines vary with orbital 
phase, and in both cases they have a skewed shape near phase 0.0 (i.e. 
at periastron). The sub-structure in the line cores near phase 0.0 is 
reminiscent of model line profiles generated by \citet{moretal2005}
for stars in binary systems with eccentric orbits. The structure in the model 
line profiles is due to flows in the outer regions of the stars that result 
from the variable tidal effects that are the 
natural consequence of an eccentric orbit. Tidal 
effects are expected to be greatest during and after periastron 
\citep[e.g.][]{moretal2010}, and so signatures of interactions might 
linger after phase 0.0.

	As a first step toward examining the general 
characteristics of the brighter star, the spectra in Figure 2 were 
shifted to the rest frame and then median-combined to form a preliminary 
template of the brighter star spectrum, and the result 
is shown in green in Figure 2. Phase-related variations 
in narrow lines that do not originate in the brighter star will 
have wavelength variations that are out of sync with the 
brighter star spectrum, and so might be suppressed by taking the median. 
Unfortunately, while broad features with motions that differ from those of 
the brighter star may be blurred when constructing 
such a template, they may not be entirely suppressed. 
Indeed, an obvious residual feature in the template spectrum is the 
broad depression blueward of the HeI 6678 line. The origin of this feature 
is discussed in the next section.

	Given the difficulties in suppressing 
features that do not belong to the brighter star, 
a proxy for its spectrum was found by comparing 
the template spectrum in Figure 2 with the spectra of a sample of bright 
stars with well-established spectral types. 
The prominent H$\alpha$ emission in the V1507 Cyg 
spectrum makes this feature of little use for spectral classification. 
Still, the wavelength region examined here contains other features that are 
sensitive probes of temperature and surface gravity. 

	A preliminary comparison with stars in the MILES compilation 
\citep[]{sanetal2006} found that the relative strengths of the SiII and 
HeI lines are consistent with a mid- to late-B spectral type. 
The narrow widths of these lines is suggestive of formation in 
a slowly rotating, low gravity environment, such as that found in the 
atmosphere of a giant or supergiant. This motivated us to examine in 
greater detail spectra of late-type B giants and supergiants 
in the 6300 -- 6800\AA\ region in the MILES archive \citep[]{sanetal2006}, 
to quantify the sensitivity of these features to temperature and surface 
gravity. We focus on the SiII lines at 6347 and 6371\AA\ , and HeI 6678, as 
these are the strongest absorption lines in our spectra.

	The ratio of the equivalent widths of SiII 6347 and SiII 6371, 
hereafter 6347/6371, is sensitive to surface gravity, in the sense that it is 
$\sim 1.3\times$ lower in supergiants than in giants of the same 
spectral type. The ratio of the sum of the equivalent widths of the two 
SiII lines to the equivalent width of HeI 6678, hereafter 
$\Sigma$SiII/HeI, is sensitive to both effective temperature and 
surface gravity. Among late B stars (i.e. B5 and later) $\Sigma$SiII/HeI 
increases as effective temperature drops, and is 
a factor of $\sim 3$ lower in supergiants than in giants of the same 
spectral class. In addition, among these supergiants HeI 6678 
is consistently deeper than either of the SiII lines, whereas the opposite 
is true among giants of the same spectral class. The SiII and HeI 
lines thus provide leverage for examining the properties of the brighter star.

	Assuming that the brighter star contributes 50 -- 60\% of the 
total light from the system at these wavelengths then 6347/6371 = 1.2, which 
is clearly consistent with a supergiant luminosity type. While measuring 
the equivalent width of HeI 6678 in the bright star template is 
complicated by the presence of the broad trough blueward of this line, 
experimentation with various stellar spectra indicate that it must have an 
equivalent width that is comparable to that of the SiII lines, which 
in turn is consistent with the supergiant classification from 6347/6371. We 
conclude based on the relative depths of the SiII lines and the 
depth of HeI6678 with respect to the Si II lines that the brighter 
star has a supergiant luminosity type.

	The DAO 1.2 meter archive 
hosted by the Canadian Astronomical Data Centre 
\footnote[1]{https://www.cadc-ccda.hia-iha.nrc-cnrc.gc.ca/en/dao/} 
was searched for B giants and supergiants that had been observed with 
the same observational set-up as described in Section 2.
Preference was given to stars that are spectroscopic standards.
The spectra recovered from the archive were reduced using the same 
procedures as those applied to the V1507 Cyg spectra. The use of stars 
observed with the same instrumental set-up and processed with the same pipeline 
reduces possible systematic errors that may affect properties such 
as line width. While software is available that parses the spectra 
of stars in binary systems, the algorithms usually assume that the light 
is stellar in origin. In the particular case of V1507 Cyg, we prefer 
using spectra of standard stars as this should 
more cleanly isolate the spectrum of the bright star given that much of 
the system light is non-stellar (see next section).

	$\eta$ UMa (B3V), $\rho$ Aur (B5V), $\chi$ Aur 
(B5I) and $\beta$ Ori (B8 I) were identified as comparison objects. 
These stars span a range of spectral types and line characteristics. 
With the exception of $\chi$ Aur these stars 
are MK spectroscopic standards \citep[]{mor1973}. 

	The V1507 Cyg spectrum is a composite of different components, 
only one of which is the brighter star, and 
experimentation found that a scaling factor of 0.6 was able to 
suppress the absorption features from the brighter star in the differenced 
spectrum. This scaling factor was found by matching 
the widths and depths of SiII6347 and 6371 as well as HeI 6678 in 
the spectrum of $\beta$ Ori (Rigel) with those 
in the spectrum of the brighter star. 
The uncertainty in the amount of light contributed by the 
brighter star at red wavelengths is estimated to be $\pm 5$\%. 

	The $\beta$ Ori spectrum is compared with that of the 
bright star template in the wavelength regions 
that sample the SiII and HeI 6678 lines in Figure 3. Also shown is 
the $\beta$ Ori spectrum scaled by a factor of 0.6 as well as
the result of subtracting the scaled spectrum from the template spectrum. 
The relative depths and widths of the SiII and HeI features are 
not well matched when the spectra of the other reference stars are used.

	While there is little ambiguity about the luminosity type 
of the brighter star, there is some uncertainty associated 
with the spectral type, as determined from $\Sigma$SiII/HeI. 
Based on comparisons (1) with the template 
stars downloaded from the DAO archive and (2) with 
stars in the MILES compilation after smoothing the V1507 Cyg spectra to 
match the 2.5\AA\ wavelength resolution of those observations,
we estimate an uncertainty of $\pm 0.5$ sub-types in the spectral class. 
Therefore, the best estimate for the spectral type of the brighter star is 
B8 I, with a possible range B7.5 I -- B8.5 I.

\begin{figure}
\figurenum{3}
\plotone{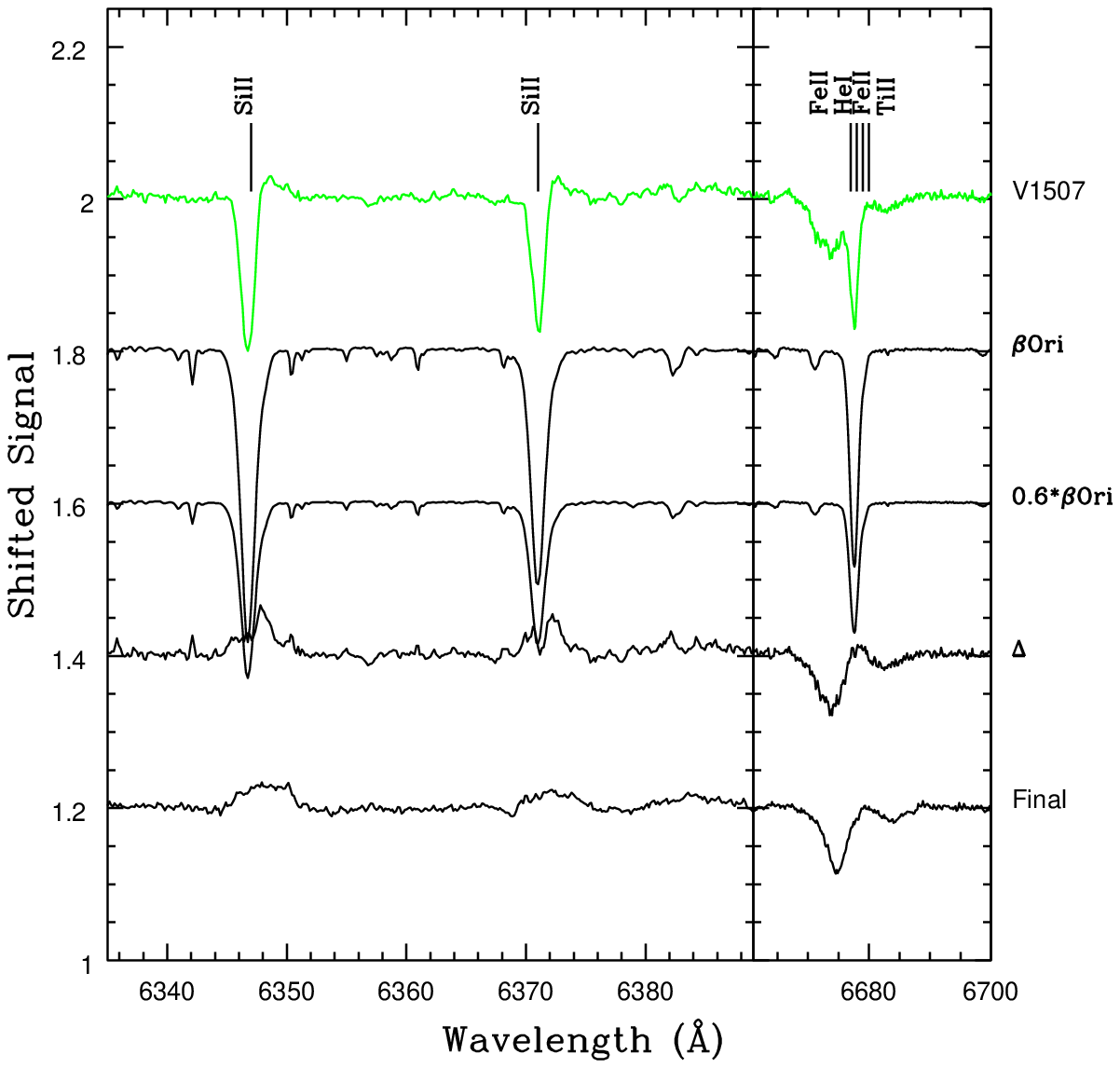}
\caption{Extracting the spectrum of the brighter star. The template 
spectrum of the brighter star that was constructed by combining the V1507 
Cyg spectra is shown at the top in green. Immediately below this is (1) the 
spectrum of the B8 I spectroscopic standard $\beta$ Ori, (2) 
the $\beta$ Ori spectrum scaled by a factor of 0.6, (3) the result of 
subtracting this scaled spectrum from the template spectrum ('$\Delta$'), 
and (4) the spectrum found by combining spectra that have had the $\beta$ Ori 
spectrum subtracted individually and then aligning the results. The bottom 
spectrum contains broad SiII emission, while HeI 6678 appears as a 
broad blue-shifted line. There is also weak emission at the rest wavelength 
of the HeI line, and a broad, shallow red-shifted component.}
\end{figure}

	To the extent that the $\beta$ Ori spectrum matches the 
spectrum of the brighter star then the radius of that star can 
be estimated with respect to Rigel. Assuming that the brighter star 
and Rigel have the same effective temperature then $\frac{L_{Bright}}
{L_{Rigel}} = \frac{R_{Bright}^2}{R_{Rigel}^2}$. 
\citet{przetal2006} find that M$_V =-7.8 \pm 0.2$ for 
Rigel. The GAIA DR3 parallax for V1507 Cyg is $1.058 \pm 0.034$ 
mas, and if the brighter star contributes 
$60\% \pm 5$ of the light to the system, for which $V \sim 7$ 
\citep[]{pavetal1979}, then $\frac{R_{Bright}}{R_{Rigel}} \sim 0.08 \pm 
0.01$. If Rigel has a radius of 74 R$_{\odot}$ with an estimated 
uncertainty of $\sim 10\%$ \citep[]{baietal2017}, then the bright star 
in V1507 Cyg has an effective radius of $\sim 5.9 \pm 0.8$R$_{\odot}$. 
Comparisons with stellar structure models and radii measurements from binary 
systems \citep[e.g.][]{toretal2010} indicates that the radius of the brighter 
star thus exceeds the ZAMS radius of a 2 -- 3 M$_{\odot}$ star (i.e. a main 
sequence star with a late B spectral type) by $2 - 3\times$. The radius 
of the brighter star is then consistent with it being an evolved object. 
 
\section{THE SPECTRUM OF CIRCUMSTELLAR MATERIAL AROUND THE COMPANION STAR}

	The spectra contain features that do not belong 
to the brighter star. The most obvious is the broad absorption trough 
blueward of HeI 6678 in Figure 2 that is seen at many orbital phases. 
This feature moves with wavelength in a manner that is 
counter to that defined by the sharp HeI line in the spectrum of the brighter 
star, suggesting that the trough is associated with the companion. 
What appear to be parts of SiII emission lines are 
also seen in the shoulders of the deep SiII absorption lines in Figure 2.

	A clear picture of the global properties of 
the HeI and SiII emission features can be gleaned 
from the bottom spectrum in Figure 3, which is the result of combining all of 
the spectra after subtracting light from the brighter star. HeI 
appears to define a P Cygni profile with a weak emission component. 
Later in this section it is shown that the strength of the emission varies 
with orbital phase, such that it is more pronounced at some points 
in the orbit. That the mean wavelength of the HeI trough is offset 
from the rest frame wavelength indicates that it 
forms in an environment that is approaching the observer. 
The offset between the bottom of the trough and the emission 
peak in the HeI profile is $4.8 \pm 0.1$ \AA\ , which corresponds to 
$216 \pm 5$ km/sec along the line of sight, while a gaussian fit to the trough 
yields a FWHM width $4.7 \pm 0.2\AA$, indicating a velocity dispersion 
of $210 \pm 10$ km/sec.

	There is also an absorption component in the HeI profile 
that is redward of the emission. The presence of such a feature suggests 
that there is material that is falling onto the envelope. 
That the redward trough is not seen at all phases suggests that the 
infalling material is not uniformly distributed, but may be 
in a stream. \citet{hut1973} also find evidence for 
infall in the profiles of some Balmer lines.

	The Si II emission lines in Figure 3 
have a flat profile. While not shown in this figure, 
weak emission from FeII is also present, and these lines have profiles that are 
similar to those of SiII. The measured width of the SiII emission lines is 
$\sim 5\AA$, which is significantly broader than the spectral resolution and
yields a characteristic rotational velocity vsini $\sim 120$ km/sec.

	The behaviour of the HeI and the 
SiII emission lines with orbital phase that remain after light from 
the brighter star is removed is examined 
in Figure 4. The phase-binned spectra shown in this 
figure were constructed by combining spectra from individual nights that 
had the contribution from the brighter star subtracted out. Offsets due to 
velocity variations within each phase bin, which can be large 
near phase 0.0, were corrected by aligning 
the deepest part of the HeI features in the residual spectra. 

\begin{figure}
\figurenum{4}
\plotone{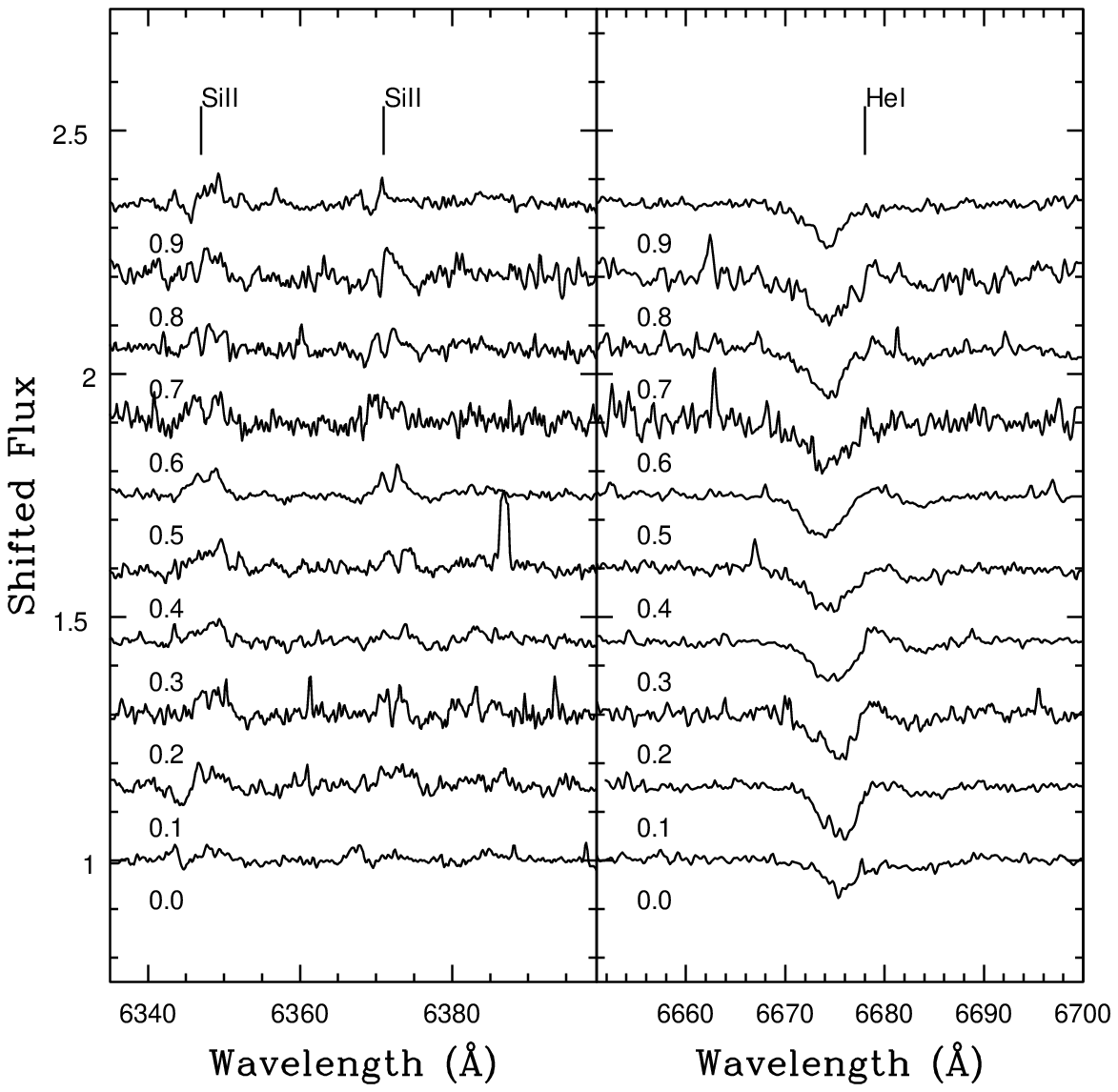}
\caption{Binned spectra with light from the 
brighter star removed. Orbital phases are listed below each spectrum. 
The central absorption in the SiII emission lines at some phases is due 
to an imperfect match with the Rigel spectrum. 
The overall strengths of the SiII emission lines vary with 
orbital phase. Absorption blueward of SiII emission at 6347\AA\ is seen at some 
phases, suggesting a P Cygni profile. Wavelengths near HeI are shown 
in the right hand panel. The broad HeI feature changes character with 
orbital phase, and the movement with wavelength is consistent with 
motions in the direction expected for the companion star. 
Broad emission redward of the HeI absorption component is also seen at most 
phases, except near periastron. The behavior of the emission lines 
suggests that the envelope around the companion is not symmetric.}
\end{figure}

	There is evidence that HeI 6678 originates in a medium that is not 
uniformly distributed around the companion, in the sense that the width and 
depth of the HeI line varies with orbital phase. The feature is narrowest and 
shallowest at phases 0.9 and 0.0 (i.e. near periastron). Weak emission is 
present redward of the trough at most -- but not all -- phases, and the 
width of the emission also varies with orbital phase. That the HeI 
emission component is not seen at all phases suggests that the emitting 
region is not symmetrically distributed around 
the companion star. If this is the case then variations 
would also be expected in the depth of the HeI trough, and these are seen.

	SiII emission is detected at most phases 
in Figure 4, although these features are noisy when considered on their own. 
The central absorption in the SiII lines that is seen at some phases is due to 
imperfect matching with the Rigel spectrum. This is not unexpected 
in a CBS with an eccentric orbit where variations in the tidal field may 
affect the outer layers of a star. There is 
also evidence of absorption blueward of the 
SiII 6347 lines at phases 0.0 and 0.1, suggesting an outward mass flow. 

	There are indications that the flow 
of material near the companion may vary from orbit-to-orbit. 
In particular, the character of features associated with the circumstellar 
environment change with time, such that orbit-to-orbit variations in the
properties of HeI 6678 occur near the same orbital phase. The detection of 
these variations was facilitated by the near-monthly cadence of the 1.2 meter 
telescope robotic observing schedule and the orbital period of the system. 

	There are three phase intervals with widths less 
than 1\% of the total orbital phase in which spectra were 
observed on three or more nights. All three sets of spectra sample the 
orbit near periastron, which is the point in the orbit where elevated 
levels of activity might be anticipated. Any changes in the character of 
the HeI feature with time in such narrow phase bins will be due to 
orbit-to-orbit differences in the absorbing/emitting medium.

	Spectra of HeI 6678 in three narrow phase 
intervals where data were recorded on three or more nights are shown in 
Figure 5. The spectra in each panel are ordered according to 
the date of observation, with spectra recorded at the earliest date 
at the bottom. Orbital phases are shown in brackets. 

\begin{figure}
\figurenum{5}
\plotone{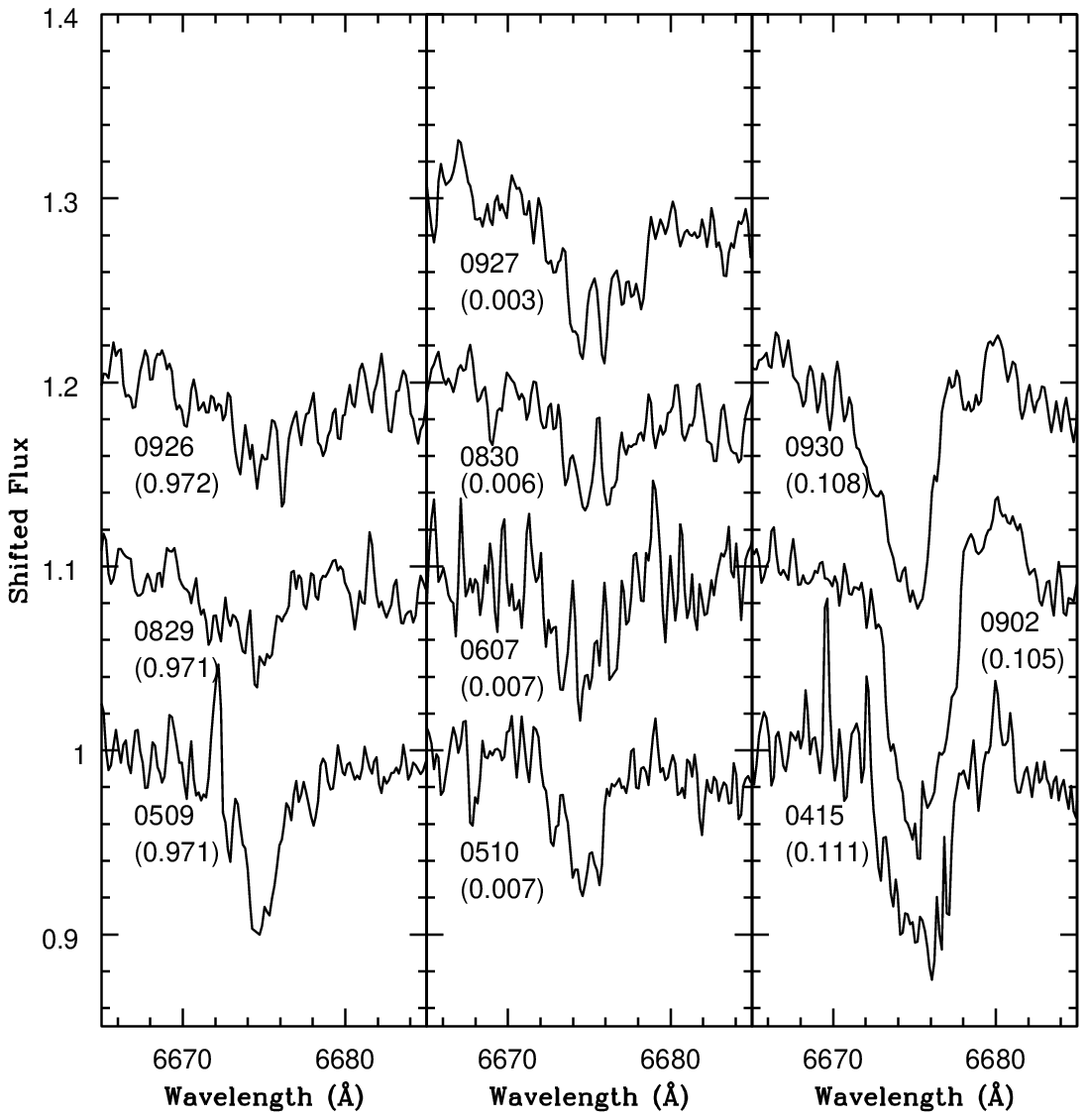}
\caption{HeI spectra that sample three orbital phase intervals near 
periastron after removing light from the brighter star. Each spectrum 
is labeled with the night in 2022 on which it was recorded, in the format 
MMDD, where MM is the month, and DD is the day of the month. The orbital 
phase of each spectrum is shown in brackets under the date of observation. 
The character of the HeI trough changes with time in all three phase intervals, 
indicating cycle-to-cycle variations in the medium in which the HeI 
trough forms. There is also evidence for line splitting near the trough center 
in the middle panel, and this is attributed to flows in the envelope 
that are driven by variations in the tidal field (see text). 
HeI emission become evident in the right hand panel.} 
\end{figure}

	There is clear evidence for a cycle-to-cycle variation in the 
depth of the HeI trough among the spectra in the left 
hand panel, which samples phases near 0.97, in the sense that the 
trough became progressively shallower from May to September 2022.
A similar, but not identical trend, is seen among the spectra near 
phase 0.005. In this case the depth of HeI in the June and August spectra 
is weaker than in May, in agreement with what is seen near phase 0.97. 
However, the HeI trough is deeper in the 0927 spectrum than in the 
0830 spectrum. This is noteworthy as the depth of the HeI trough in the 
spectrum recorded on the previous night (0926, phase 0.972) is similar to that 
recorded on 0829. This suggests that significant changes in the 
behaviour of the HeI trough can occur over the time span of a day. 

	Some of the spectra in the middle panel of Figure 5 have 
sub-structure in the trough profile, in the sense that there are two components 
present near the center of the trough. The 0927 spectrum (phase = 0.003) 
has the highest S/N ratio of the four spectra in the 
middle panel, with an estimated noise of $\pm 1\%$ in the normalized flux. 
There is a spike near the center of the HeI trough in the 0927 spectrum 
that has an amplitude of $\sim 5\%$, and so is well above the noise 
level. A central spike with a similar wavelength and amplitude 
is also seen in the 0830 spectrum (phase = 0.006), although the noise in that 
spectrum is slightly higher than in the 0927 spectrum. Hence, structure in 
the He trough is seen near periastron in at least some of the spectra.

	The sub-structure in the central regions of the He trough 
is reminiscent of behaviour seen in the model spectra of stars
on eccentric orbits at orbital phases near periastron 
that are discussed by \citet{moretal2005}. The models attribute this 
sub-structure to flows induced by the tidal field as the stars move 
in eccentric orbits. To the extent that these models apply to V1507 
Cyg then the structure in the He trough suggests that the envelope 
that produces this feature might contain such tidally-driven flows.
It should be recalled that similar sub-structure is seen in the centers of 
the SiII lines and HeI lines in the spectrum of the brighter star at 
similar phases (e.g. Figure 2). The difference between the centroids in the 
HeI lines in the 0927 and 0830 spectra in the middle panel of Figure 5 is 
$\sim 1.5$\AA\ , which corresponds to a velocity difference of 
70 km/sec along the line of sight.

	The HeI trough is much deeper in the three 
spectra near phase 0.105 than in the other two phase intervals. 
In the context of an asymmetric envelope geometry this is consistent 
with a thicker part of the envelope being seen 
along the line of sight at this phase. The HeI trough 
is deepest in the 0902 spectrum. HeI emission is also present in this phase 
interval, and the strength of the emission appears not to change with 
time. The radial velocity curve of the primary star is steep near periastron, 
and the motion of the deepest part of the HeI trough 
from phase 0.97 to 0.11 is qualitatively consistent 
with motion in the same direction as that of the brighter star, as expected if 
motions in the envelope are coupled with the orbit of the brighter star.

\section{RADIAL VELOCITIES AND MASS ESTIMATES}

\subsection{Radial Velocities of the Brighter Star}

	Radial velocities of the brighter star were measured 
by cross-correlating the system spectra with the spectrum of Rigel 
in the wavelength interval 6335 - 6465\AA\ , and the results are shown as 
black crosses in Figure 6. The left hand panel shows velocities measured 
as a function of time, while the right hand panel shows a phased
radial velocity curve constructed using the \citet{hut1973} ephemeris. 
Radial velocity measurements that are the average of 
the HeI and metal values in Table 1 of \citet{mer1949}, phased 
using the \citet{hut1973} ephemeris, are also plotted in the right hand panel.

\begin{figure}
\figurenum{6}
\plotone{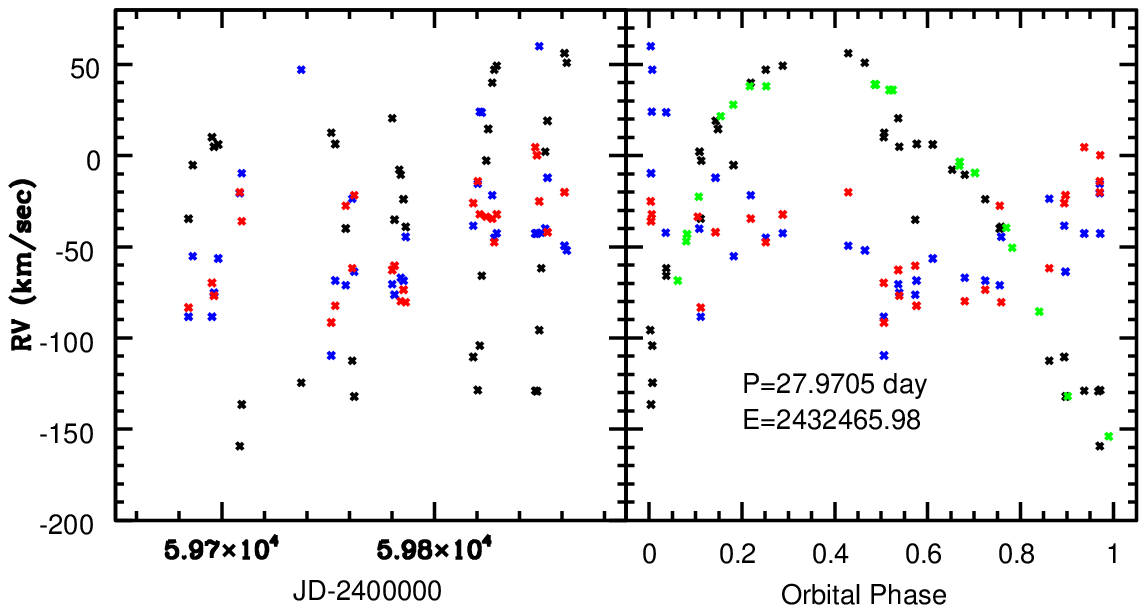}
\caption{(Left hand panel:) Radial velocities. Velocities of the brighter star 
are shown as black crosses, while those obtained from the envelope 
spectra are shown as blue (HeI) and red (SiII emission lines) crosses. 
(Right hand panel:) Velocities from the left hand panel phased using 
the ephemeris found by \citet{hut1973}. The green squares 
are the means of the velocities for HeI and metal lines listed in Table 1 of 
\citet{mer1949} with phases computed from the 
\cite{hut1973} ephemeris. The velocities for the 
bright star measured from the DAO spectra agree with the Merrill 
measurements, indicating that the orbital properties of the 
system have then been stable to within $\pm 10^{-3}$\% over the time 
span of a century. That the amplitude of the radial velocity curve of 
the bright star is larger than the range in radial velocities 
obtained from the envelope spectra indicate that the 
brighter star is less massive than the companion.}
\end{figure}

	The measurements made by Merrill track those obtained from the DAO 
spectra to within 1 - 2\% of an orbital cycle. This indicates that the 
ephemeris has not changed greatly: the orbital characteristics of the brighter 
star have remained more-or-less stable over a time span in excess of roughly 
one century. A dispersion of 1 - 2\% of an orbital cycle over this timeline 
indicates that the period has remained stable to within $\pm 10^{-3}\%$.

\subsection{Radial Velocities of the Envelope}

	In Section 4 it was noted that the features in the spectra that remain 
after removing the light from the bright star move with wavelength in a manner 
that is opposite to that of the brighter star, suggesting an association with 
the companion. To the extent that these features 
track the motion of the companion then they allow the masses 
of both objects in the system to be determined. Two wavelength regions 
in the envelope spectra were examined to measure radial velocities. One was in 
the wavelength interval containing the SiII emission lines, while the other 
covered HeI in the wavelength interval 6650 - 6700\AA\ . 
Velocities were obtained by cross-correlating the residual spectra in these 
wavelength intervals with the median template shown in Figure 3.

	The velocities measured from the residual spectra are 
plotted in Figure 6. The mean difference between the HeI 
and SiII emission line velocities is 4 km/sec, with a standard deviation 
of $\pm 25$ km/sec, thereby suggesting that they track a common source. The 
dispersion between the two sets of measurements is greatest near periastron.

	That the velocities obtained from the envelope spectra do not define a 
tight velocity curve in the right hand panel of Figure 6 is not due to poor 
data quality. In the previous Section it was noted that there are 
cycle-to-cycle variations in the character of HeI near periastron, and this 
is where the difference between the HeI and SiII velocities is largest. 
Individual HeI envelope spectra near phase 0.0, where there is a 
large dispersion in HeI-based velocities, are shown in Figure 5. The 
two highest HeI envelope velocities were recorded on the nights of 
0830 and 0831, and the spectra of those nights do not 
have markedly poor signal-to-noise ratios. Still, the 0830 
spectrum, shown in the middle panel of Figure 5, has 
the shallowest HeI trough in this phase interval. 
This suggests that the velocities measured from the envelope spectra 
are likely susceptible to activity associated with this point in the orbit. 
Spectra that span a number of orbital cycles are then desirable to average 
out the affect of this activity. This is a source of uncertainty when 
calculating masses for this system.

\subsection{Mass Estimates}

	Given (1) the scatter in the velocities 
obtained from the envelope features and (2) that the velocity curve for 
the brighter star is very well-defined, so that the orbital elements 
deduced from those velocities are well-determined, then we have opted 
not to obtain an independent orbital solution from the envelope 
velocities. Instead, it is assumed that (1) the 
eccentricity of the companion orbit is the same as that of the brighter star, 
and (2) the arguments of periastron ($\omega$) for the two stars 
differ by $180^o$. This simplifies the task of estimating a mass ratio. 

	While the velocity measurements made from the envelope spectra 
may not define a clear-cut radial velocity curve, the majority of 
them still contain information about the orbit of the companion. 
Evidence to support this claim comes from a comparison of the velocities 
of the bright star and those obtained from the envelope spectrum. 
The relationship between the velocities of the bright star and those 
obtained from the envelope is examined in Figure 7, where the velocity 
measurements obtained from the SiII absorption lines in the spectrum of the 
bright star are compared with velocities obtained from the HeI 
profile (top panel) and the SiII emission lines (middle panel). 
The lower panel compares the mean of the HeI 
and SiII emission velocities with those of the bright star.

\begin{figure}
\figurenum{7}
\plotone{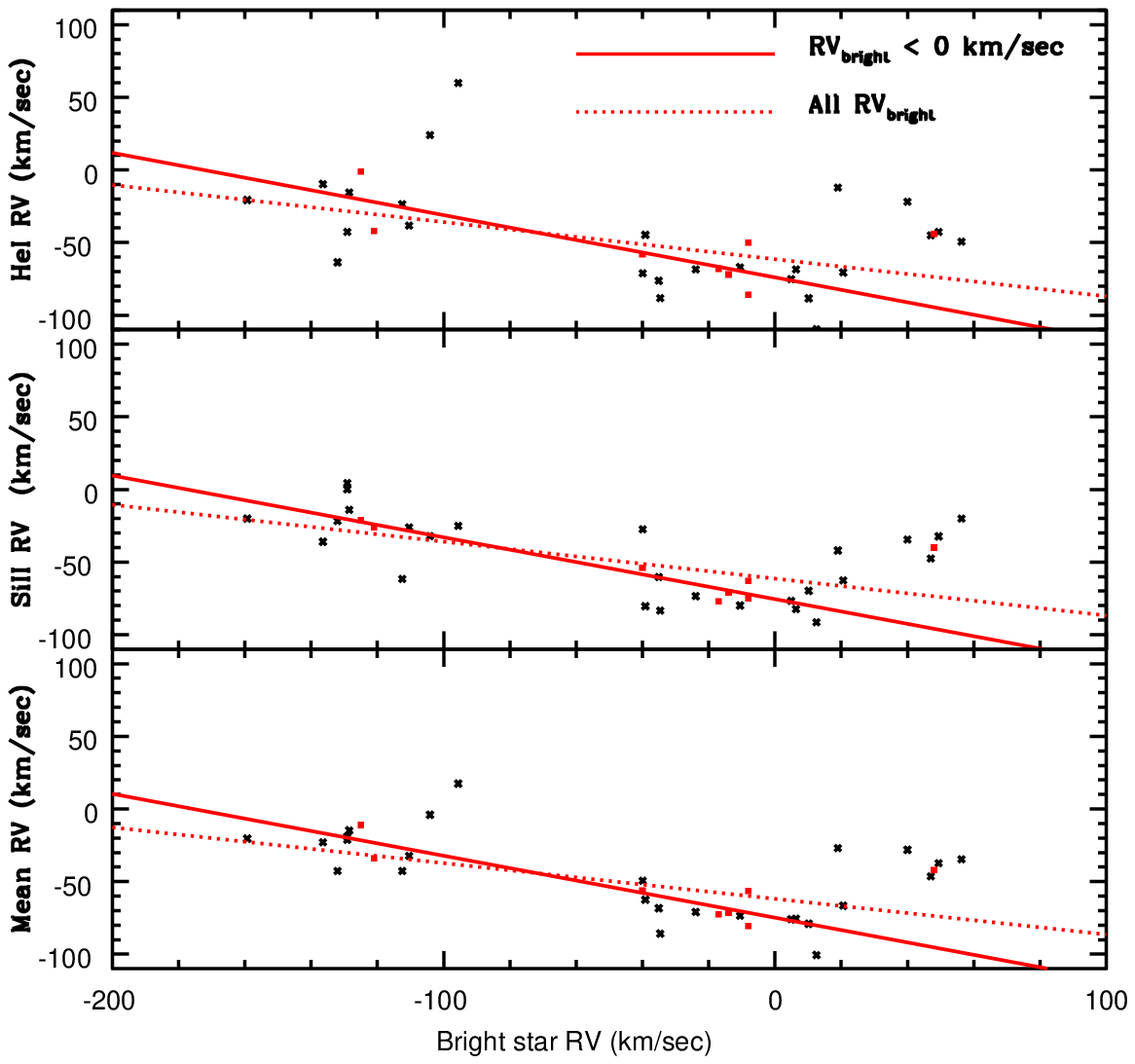}
\caption{Radial velocity measurements of the bright star are compared with 
velocities measured from the HeI profile (upper panel) and SiII 
emission (middle panel) in the envelope spectra. The mean of the combined 
HeI and SiII emission velocities is shown in the bottom panel. The 
filled red squares are the mean velocities in phase intervals 
that are 0.1 units wide where at least two velocity measurements were made. 
The dashed red lines show a least squares fit to all of the phase-binned 
velocities, while the solid red lines show a least squares fit that does not 
include the outlying point in the binned measurements that 
has v$_{brightstar} > 0$ km/sec. That the velocities measured 
from the HeI profile and the SiII emission lines are 
anti-correlated with the velocities measured for the bright star 
over much of the orbital velocity range of the bright star is consistent with 
them tracking the orbital motion of the companion.}
\end{figure}

	If the velocities obtained from the envelope spectra track the 
motion of the companion then, given the assumptions about the orbit of the 
companion discussed above, the velocities of the bright star and those 
measured from the HeI and Si II emission lines should be anti-correlated. 
The slope of the relation between the two then provides a statistical means of 
determining the relative radial velocity amplitudes of the bright star and its 
companion. In fact, there is a clear tendency for the velocities
of the brighter star and those made from the SiII 
emission and the HeI profile to be anti-correlated over 
much of the velocity range covered by the brighter star. This is to be expected 
if the velocities measured from the HeI profile and SiII emission track 
the motion of the companion and its envelope over much of the orbital cycle. 
However, there is considerable scatter and degradation of the relation 
from linearity when $v_{bright} > 0$ km/sec, which corresponds to 
phases 0.1 -- 0.4.

	What is the cause of the outlier points in Figure 7 at 
post-periastron phases? The radial velocity curve 
of the brighter star at these phases is not 
distorted, indicating that the cause of the outlying points is associated 
with the envelope around the companion. The HeI 6678 profile indicates that 
the envelope is extended and contains an expanding component, and so is 
susceptible to tidal distortions that are amplified near periastron. 
Moreover, the variations in the HeI profile with phase that are seen 
in Figure 4 indicate that the envelope is not uniformly distributed 
around the companion. We thus suspect that the outlier points in the envelope 
velocities in Figure 7 are the result of flows within the envelope that are 
induced by the variable tidal effects that result from the 
eccentric orbit and drive asymmetries in the envelope.
These flows cause the velocities measured from the envelope spectrum 
immediately after periastron to depart from those that track the orbital 
trajectory of the companion. 

	Given the dispersion in the envelope velocities, mean 
velocities were calculated in 10\% wide phase intervals, and 
these are shown as the solid red squares in Figure 7. It can be seen 
that taking the mean suppresses some of the scatter in Figure 7. 
Least squares fits were then made to the binned velocities, 
and the results are shown with the dashed and solid red lines in that figure. 
A least squares fit is appropriate given that the uncertainties in the 
velocities of the brighter star are much smaller than those of 
the other source, as is evident from the modest scatter among 
the velocities obtained from the bright star spectrum in Figure 6. 

	If the SiII emission and HeI features track the 
motion of the companion star then the slope of the relationships in 
Figure 7 is the ratio of the velocities of the two stars. 
In fact, the slopes defined by the HeI and SiII emission measurements 
are identical within their estimated uncertainties. 
However, given the smaller scatter, the SiII emission 
line measurements are adopted here to estimate K$_2$, and these yield 
$\frac{\Delta V_{envelope}}{\Delta v_{brightstar}} = 0.24 \pm 0.12$ 
(all phase-binned points), and $\frac{\Delta v_{envelope}}
{\Delta v_{brightstar}} = -0.42 \pm 0.05$ (not including the 
mean for phase 0.3, which corresponds to $v_{brightstar} > 0$ km/sec). 
We favor the latter slope given that it has a much 
smaller uncertainty, and so provides a better match to 
most of the envelope velocities. In either case the ratio of the 
component velocities indicates that the companion is 
more massive than the brighter star, and so mass transfer has 
progressed to the point where the mass ratio has been reversed -- what was 
once the more massive star is now the less massive star. 

	The well-defined radial velocity curve for the 
brighter star indicates that K$_1 = 105.2 \pm 1.2$ km/sec \citep[]{hut1973}. 
The slope of the relationship in the middle panel of Figure 7
indicates that K$_2 = (0.42 \pm 0.05) \times K_1$, so that 
K$_2 = 44.2 \pm 5.3$ km/sec. With K$_1$ and K$_2$ known then masses 
can be estimated. If $e = 0.388$ and the period is 27.9705 days 
\citep{hut1973}, then M$_1$sin$^3$i = 2.44$\pm 0.32$M$_{\odot}$ 
and M$_2$sin$^3$i = 5.33$\pm 0.29$M$_{\odot}$.

	An estimate of the orbital inclination is required 
to compute final masses, and we consider two possibilities. 
First, \citet{ber1998} estimate that 
$i = 46.4 \pm 2.1$ degrees, such that M$_1 = 6.4 \pm 
0.9$M$_{\odot}$ and M$_2 = 14.0 \pm 0.9$M$_{\odot}$. We consider this 
to be the best estimate of the system masses. The total system mass at the 
present day is then in excess of 20M$_{\odot}$, which far exceeds the upper 
mass limit estimated by \citet{dav2023} based on the properties of stars 
thought to be in a moving group with V1507 Cyg. 

	The second possibility assumes an upper limit 
for the inclination, based on the absence of 
eclipses in the light curve. The use of an upper limit for the 
inclination results in a lower limit for the masses. Adopting an upper 
limit of $60^o$ for the inclination then we find that M$_1 \geq 3.78 
\pm 0.49$M$_{\odot}$, and M$_2 \geq 8.21 \pm 0.45$M$_{\odot}$. 
We further note that if the slope of the dashed line in 
the middle panel of Figure 7 is adopted then K$_2 = 25.2 \pm 12.6$ km/sec.
If this value of K$_2$ is adopted then the mass of the brighter star drops 
by $\sim 0.4\times$, while that of the companion drops by $\sim 0.8\times$.

	The mass estimates for the brighter star presented above suggest 
that it is significantly less massive than $\beta$ Ori, which 
\citet{przetal2006} estimate to have a mass $21 \pm 
3$M$_{\odot}$ from comparisons with evolutionary tracks. 
The relative masses of $\beta$ Ori and the brighter star are consistent 
with the latter having been stripped by mass transfer. As for the companion 
star, an unobscured main sequence star with the mass estimated above 
would have M$_V \sim$ --3 to --4, and so will be intrinsically 
brighter than the B8 I star. However, such a star might not be seen 
at visible wavelengths if it is masked by an optically thick accretion disk 
or envelope.

\section{H$\alpha$}

	There are a number of potential sources of H$\alpha$ emission in 
CBSs, including an accretion disk, hot spots, jets, and circumstellar/system 
envelopes. Emission from these (and other) sources may combine to produce 
complex emission line structures. What might appear initially as a single 
broad H$\alpha$ emission feature at low or moderate wavelength 
resolution may actually be a blend of distinct 
features from multiple components \citep[e.g. Figure 16 of][]{broetal2021}. 

	H$\alpha$ profiles extracted from the phase-binned spectra 
are shown in Figure 8. These spectra have had the contribution 
from the brighter star, which is the H$\alpha$ line in the 
$\beta$ Ori spectrum scaled by a factor of 0.6, subtracted out. The scaled 
$\beta$ Ori line is small when compared with the emission 
line in the system spectrum, and so its removal has only a minor effect 
on the overall properties of the H$\alpha$ emission from the system.

\begin{figure}
\figurenum{8}
\plotone{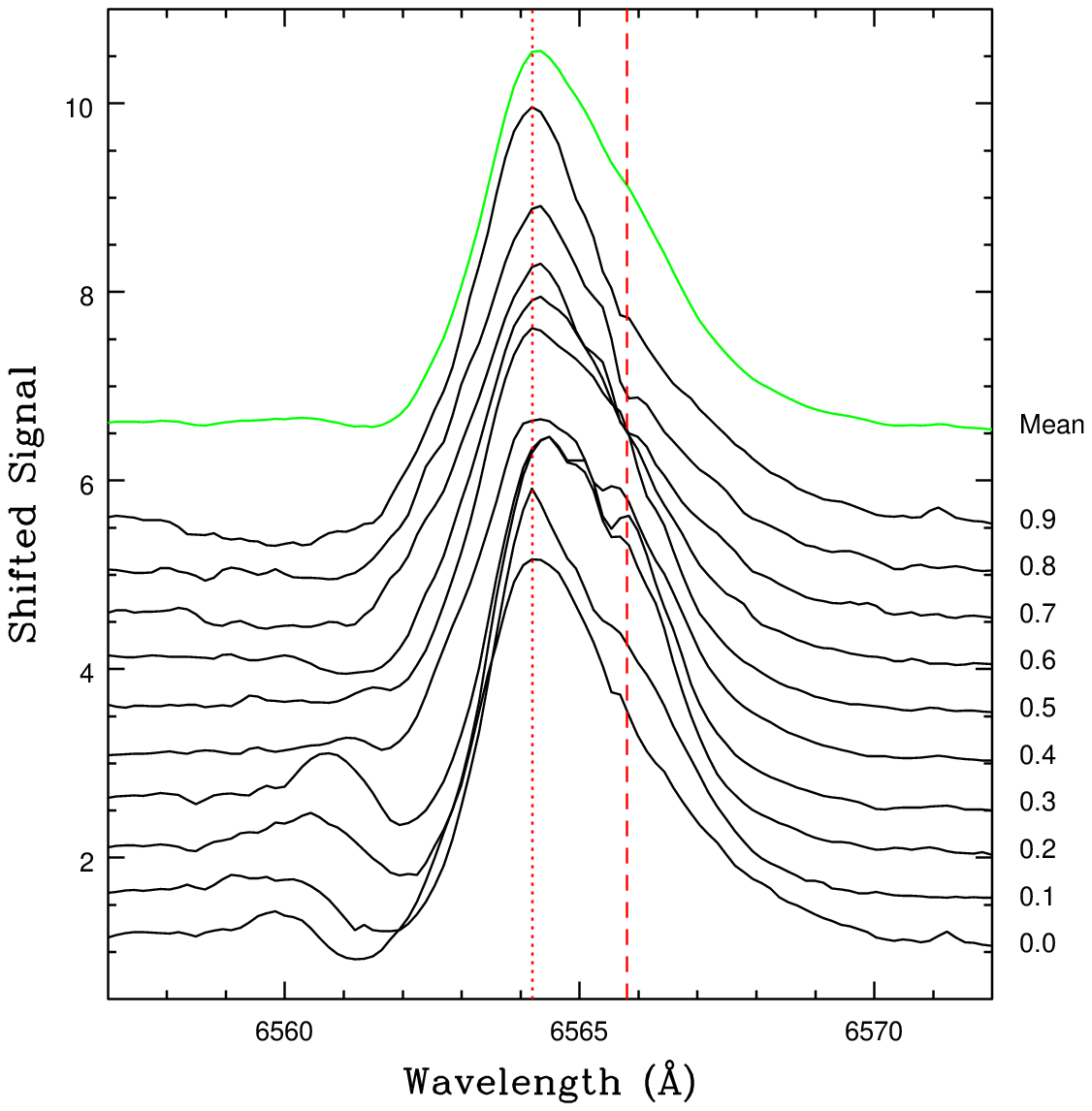}
\caption{H$\alpha$ profiles in phase-binned spectra. The contribution from 
the brighter star, which is assumed to have H$\alpha$ 
in absorption based on the template $\beta$ Ori spectrum, has been 
removed. The mean of the profiles is shown in green. 
The overall morphology of the H$\alpha$ profile at 
some phases is similar to what is seen in the spectrum 
of some Herbig AeBe stars \citep[e.g.][]{ham1992}. The vertical red 
lines mark the approximate locations of the peak in H$\alpha$ emission 
(dotted line) and a second bump that is slightly redward of the 
main peak that is most clearly seen between phases 0.2 and 0.5 (dashed line). 
The absorption trough between 6559 and 6560\AA\ that is present 
between phases 0.0 and 0.4 is likely due to blueshifted H$\alpha$ absorption.}
\end{figure}

	The width of the H$\alpha$ profile changes with orbital phase. 
At phases 0.4 and 0.5 the FWHM of the profile is $3.4 \pm 0.1\AA$ , 
while at other phases it is $\sim 3.0 \pm 0.1\AA$.
The broadening of the profile at intermediate phases is due to a 
secondary peak, which we will refer to as the 'red bump'. This feature 
is $\sim 1.5$\AA\ redward of the main peak and 
is most pronounced near phases 0.3 -- 0.4. 

	There is a trough blueward of the dominant emission 
feature at phases 0.0 -- 0.4 that is likely due to 
blueshifted H$\alpha$ absorption. Thus, it likely forms in an 
expanding envelope. This trough propogates to 
progressively longer wavelengths from phases 0.0 to 0.4, and 
disappears from phase 0.5 onwards. If the medium in which the trough 
formed was symmetrically distributed and moving 
with the brighter star then the trough should reappear at 
phases 0.8 and 0.9, which is not the case. Hence, we suspect that 
this feature is associated with an ouflow of material that is to one side 
of the brighter star.

	Additional insights into the behaviour of H$\alpha$ can be gleaned 
by studying departures from a reference profile. To this end, 
a mean H$\alpha$ profile was constructed from the spectra in 
Figure 8, and the result is shown as the green line at the top of that figure. 
The results of subtracting this mean profile from each of the 
phase-binned spectra are shown in Figure 9, and the 
residuals show clear trends with phase. 

\begin{figure}
\figurenum{9}
\plotone{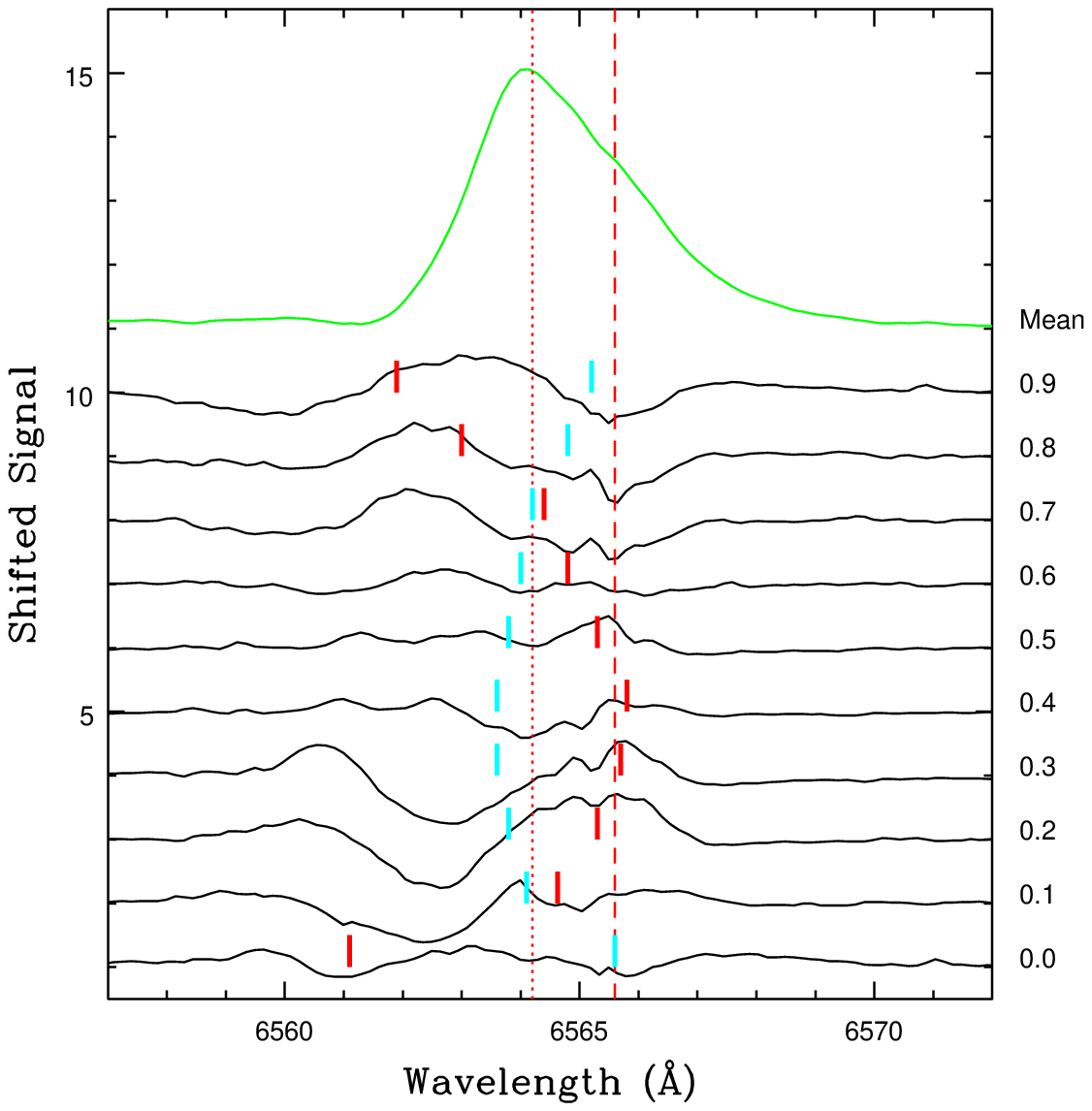}
\caption{Residual spectra that are the result of subtracting 
the mean H$\alpha$ profile (shown in green) from the phase-binned 
spectra in Figure 8. The vertical dotted and dashed red lines mark the 
same features as in Figure 8. The red hash marks above each curve indicate 
the expected location of H$\alpha$ at that phase if it followed the radial 
velocity curve of the brighter star. The corresponding cyan hash marks 
indicate the expected location of H$\alpha$ based on the radial velocities 
of the companion, as inferred from the solid line in the 
middle panel of Figure 7. The peak in the residuals 
propogates in a manner that is consistent with the motion 
of the brighter star, as expected if the bulk of the emission is 
associated with that star.}
\end{figure}

	A systematic propogation of structure 
in the residuals with orbital phase can be 
seen in Figure 9. There is a peak in the 
residuals near 6563\AA\ at phase 0.9, while at phase 0.2 the peak 
is centered near 6565\AA. The expected location of H$\alpha$ with 
wavelength based on the radial velocity curve of the brighter star 
is marked for each residual curve, and it can be seen that the 
peak in the residuals more-or-less propogates with phase in 
a manner that is consistent with the motion of the brighter 
star. While the agreement between the peaks in the residual curves and the 
expected location of H$\alpha$ is not perfect, the results in Figure 9 
suggests that the bulk of the H$\alpha$ emission 
is associated with the brighter star and its environs.

	There are orbit-to-orbit changes in the H$\alpha$ 
profile over time scales of a few weeks. 
Evidence for this is presented in Figure 10, where 
differences in H$\alpha$ characteristics within the 
narrow phase intervals that were considered in Figure 5  
are examined. It should be recalled that these 
spectra examine the system shortly before, during, and after periastron. 

\begin{figure}
\figurenum{10}
\plotone{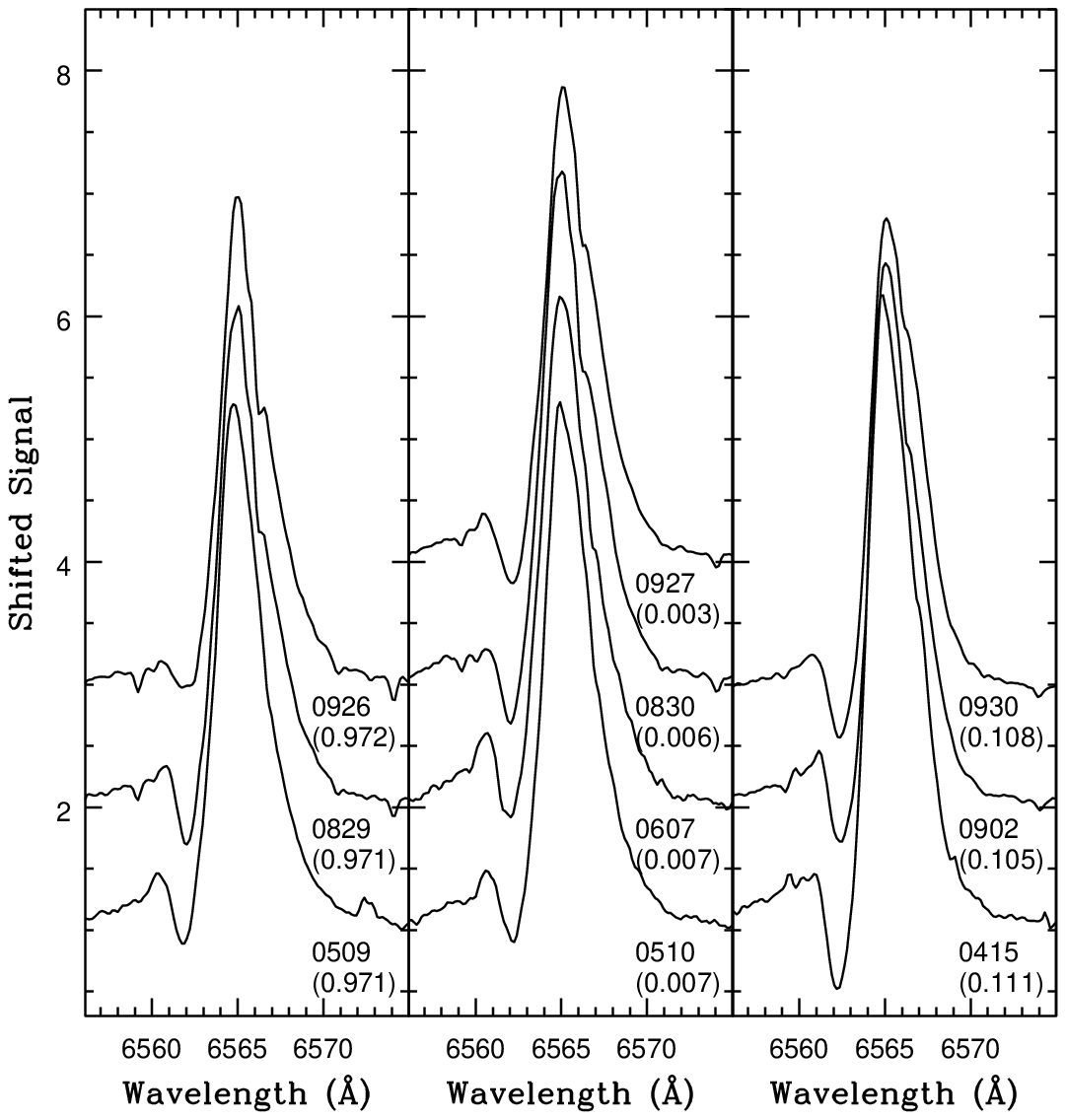}
\caption{Same as Figure 5, but showing H$\alpha$. 
The overall shape of H$\alpha$ emission varies with time 
in each phase interval in a more-or-less consistent manner, in the sense that 
the red bump appears in spectra recorded after June 
2022. The depth of the trough to the left of the main emission feature 
weakens with the appearance of the red bump, suggesting that they might be 
related. The relative height of H$\alpha$ emission is roughly 
constant with time in the left hand panel, and changes only slightly 
with time in the middle panel. However, significant changes 
in the height of H$\alpha$ occurs with time in the right hand panel, in the 
sense that the peak height of the emission weakens with time.
This is accompanied by a decrease in depth of the blue absorption trough.}
\end{figure} 

	It is apparent from the left hand and central panels that 
the red bump becomes pronounced after June 2022. This is also seen 
in the right hand panel, where the red bump is more pronounced in the 
September spectra when compared with the April spectrum. 
The asymmetric shape of the H$\alpha$ profile in the spectra 
recorded from April to June suggest that the red bump is also present 
at that time, but is blended with the main H$\alpha$ emission feature. 

	There are also noticeable changes in the overall height of H$\alpha$ 
among the spectra in the right hand panel (i.e. a few days after 
periastron). These changes are in the sense that the relative strength 
of H$\alpha$ emission weakens with time. These changes are accompanied by a 
change in the depth of the absorption trough.

	The impact of periastron on the line emission in April 2022 differs 
from that in September 2022. This might reflect orbit-to-orbit 
changes in the envelope around the brighter star. 
The spectra in the right most panel of Figure 10 
sample phases near $\sim 0.1$, and models examined by \citet{moretal2010}
suggest that the impact of periastron passage is most visible following 
periastron.

	\citet{hut1973} describe variations in the 
characteristics of H$\alpha$ that occur over time scales of 10 minutes. 
Variations on such a short timescale might be surprising 
given the consistency in the overall strength of H$\alpha$ 
when averaged over the $\sim 20$ minute timescales 
that are covered by our nightly spectra. While typically only three 
spectra of V1507 Cyg were recorded on any given night for this program, there 
were nights during which five 300 sec exposures were recorded, 
and the spectra from those nights were searched for evidence of 
flickering in the strength of H$\alpha$. No evidence for changes in the 
characteristics of H$\alpha$ were found over the $\sim 30$ minute time 
interval covered by these exposures.

\section{THE INTERSTELLAR MEDIUM NEAR V1507 CYG}

	The ISM near V1507 Cyg may provide clues 
into its recent past. Assuming conservative mass transfer (i.e. 
the present-day total system mass is the same as the total mass at the time of 
formation) then the mass measurements made in Section 5 indicate that the 
initially more massive star formed with a mass $\gtrapprox 10$M$_{\odot}$. 
The system then has an age that is $\lessapprox 25$ Myr. 
Given this relatively young age then if V1507 Cyg formed in a 
cluster or an association with other massive objects then signatures in the 
surrounding ISM such as a supernova remnant (SNRs) might be present. 
We have thus searched for evidence for structures in the ISM near 
V1507 Cyg using the CGPS, which is a survey that mapped a large fraction 
of the northern Milky-Way 1420 Mhz ($\lambda = 21$ cm) and 
408 Mhz ($\lambda = 74$ cm).

	Images of the area around V1507 Cyg were downloaded from the CADC 
website \footnote[1]{https://www.cadc-ccda.hia-iha.nrc-cnrc.gc.ca/en/cgps/}, 
and the results are shown in Figure 11 . The intrinsic angular resolution 
of the maps is $\sim 1$ arcmin at 1420 MHz and a few arcmin at 408 MHz. 
The 1420 MHz image in Figure 11 has been smoothed 
with a $\sigma = 2.4$ arcmin Gaussian to match the 
angular resolution of the 408 MHz image.

\begin{figure}
\figurenum{11}
\plotone{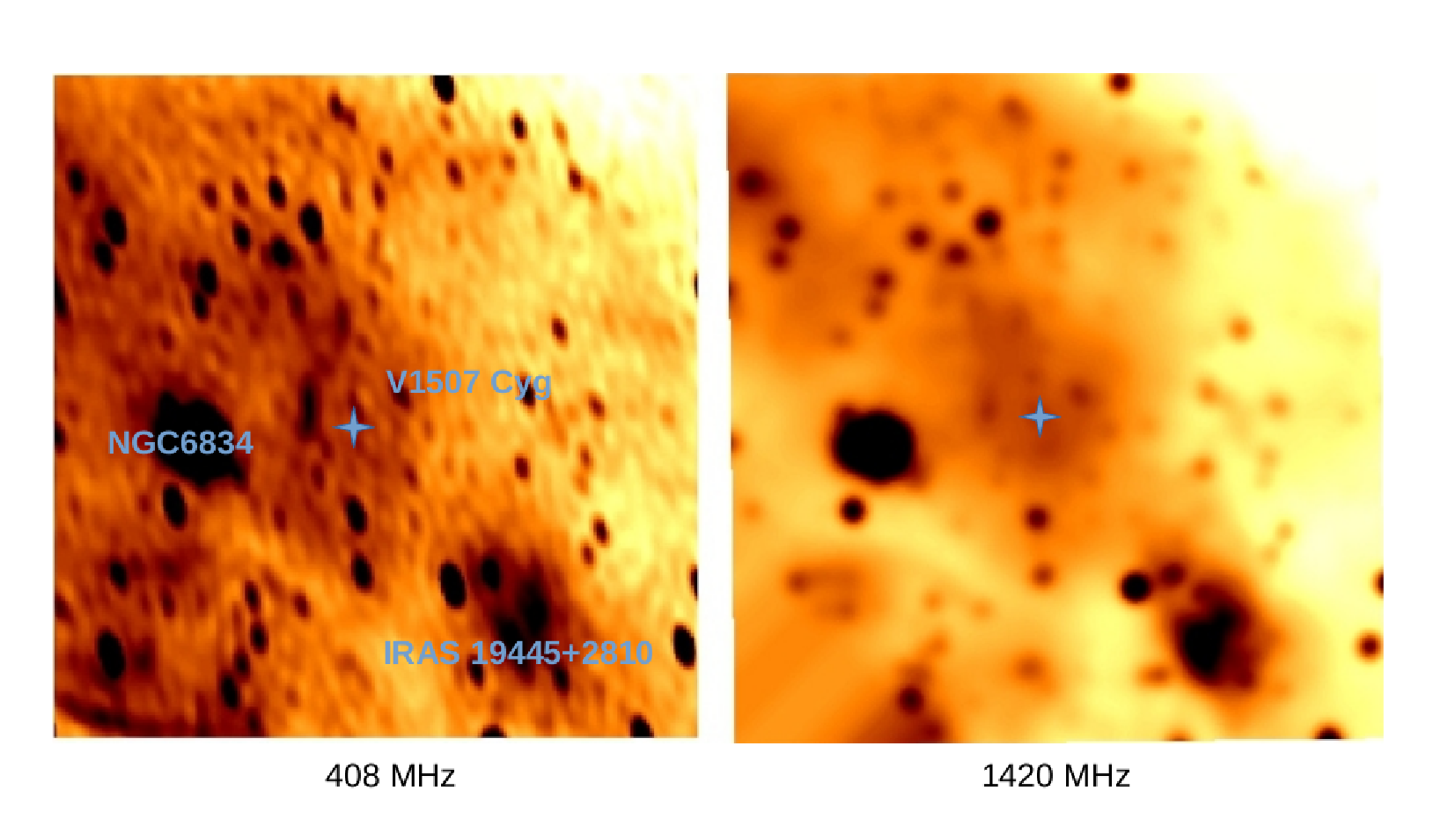}
\caption{Maps of the area around V1507 Cyg at 408 Mhz (left panel) 
and 1420 MHz (right panel). Both were recorded as part of the CGPS. 
The 1420 MHz map has been smoothed to match the angular resolution of the 
408 MHz map. Each map covers $180 \times 180$ arcmin, with North at the top 
and East to the left. V1507 Cyg, marked by the blue star, is at or near the 
center of a diffuse emission structure that is most obvious in the 
1420 MHz map. The intermediate age cluster NGC 6834 and the diffuse source 
IRAS 19445+2810, discussed in the text, are also labeled. Neither of these 
appear to be associated with the HI emission that is centered on V1507 Cyg.}
\end{figure}

	Figure 11 reveals a mix of resolved and 
unresolved sources on the sky close to V1507 Cyg. 
A complex distribution of objects at these wavelengths
is not unexpected as the line of sight at low Galactic latitudes 
passes through the disk. Not only is the disk an area 
that contains a rich population of objects that might form structures in 
the ISM, but sources that originate over a large 
range of distances are also sampled. Sources that are physically unrelated 
may then be superimposed in Figure 11, introducing potential ambiguity in 
source identification; a source close to a target of interest on the sky may 
actually be in the foreground or background. Keeping this caveat in mind, we 
note that V1507 Cyg is at the center of diffuse emission at 408 MHz 
and 1420 MHz that has a more-or-less circular morphology. 

	Young stars are located close to V1507 Cyg on the sky, 
although their parallaxes indicate that these are 
background objects \citep{dav2023}. Could these be 
responsible for the emission centered on V1507 Cyg in Figure 11? 
To answer this question we briefly discuss two young objects that are 
viewed close to V1507 Cyg.

	The 80 Myr old cluster NGC 6834 contains a number of emission line 
stars \citep{pauetal2006}, and it is the brightest single source in Figure 11. 
This cluster is located within 50 arcmin of V1507 Cyg, 
and is roughly midway between V1507 Cyg and the 
left hand edge of Figure 11. It is clearly displaced from the 
HI structure that is centered on V1507 Cyg. 

	Another bright source is IRAS 19445+2810, which is the diffuse extended 
collection of objects to the north west of V1507 Cyg. This source does not 
appear in W1 and W2 images in the WISE survey, leading us to speculate that it 
is a heavily obscured -- possibly massive -- young cluster. However, like 
NGC 6834, IRAS 19445+2810 does not appear to be related to the HI emission 
centered on V1507 Cyg. Rather, it appears to be surrounded by its own diffuse 
HI halo.

	There are other moderately bright sources that are close 
to V1507 Cyg in Figure 11. These appear to be point sources at the 
angular resolution of the maps, and are offset from the 
center of the emission. In summary, there remains ambiguity in the 
line-of-sight distances to V1507 Cyg and the emission that surrounds it. 
Still, that V1507 Cyg is at the center of an emission feature indicates 
that a more detailed investigation of the ISM around it may prove to be 
rewarding.

\section{DISCUSSION \& SUMMARY}

	Spectra that were recorded over a six month period in 
2022 have been used to examine the properties of the CBS V1507 Cyg. 
Previous studies have considered this system in the context of 
a single line spectroscopic binary, and the mass function is suggestive of a 
massive -- but faint at visible wavelengths -- companion. The new spectra 
sample 6 complete orbital cycles and have a common wavelength 
coverage of $0.63 - 0.68\mu$m with a wavelength resolution 
$\frac{\lambda}{\Delta \lambda} \sim 17000$. Prominent absorption 
features that can be attributed to the brighter star are present. 
While the spectra do not appear to contain features that originate 
directly in the photosphere of the companion, 
features that track the motion of this object 
have been identified, allowing new insights to be 
gained into the nature of this system.

\subsection{System Components: Evidence of An Evolved Binary System}

\subsubsection{The brightest star}

	The deepest absorption lines in the DAO spectra are SiII 6347, SiII 
6371, and HeI 6678. The SiII 6347 doublet is a resonance transition, and there 
is no evidence for an interstellar contribution, such as might be 
expected if there were an optically thick circumsystem envelope. 
These deep absorption lines track the motion of the 
brightest star, and the measured velocities are in 
excellent agreement with those obtained by \citet{mer1949}, who examined 
spectra that were recorded over more than two decades. The amplitude of 
the radial velocity curve defined by the brighter star is larger than the 
range in velocities measured from the envelope spectrum, indicating that 
the brighter star is less massive than its companion.

	Models predict that the structural properties of the donor star will 
be affected by tidal interactions, and these can have an impact on features 
in the spectrum. \citet{moretal2005} model line profiles in stars 
with eccentric orbits, and find that flows induced 
in a star during periastron passage can alter line profiles. 
The predicted line profiles are similar to those seen in the 
SiII and HeI lines in the spectrum of the brighter star near periastron.

	That the orbital properties of the system have not changed 
over roughly a century is consistent with a system in which there is not a 
high rate of mass exchange at the present day. 
A low rate of mass transfer is consistent with the absence of a 
thick circumsystem envelope. Such an envelope is expected to form 
during the rapid phase of mass transfer when the more massive 
star expands to fill its Roche lobe \citep[e.g.][]{desetal2015}. 

	The properties of the brighter star are consistent with it being 
in a binary system that has undergone extensive mass transfer. 
Comparisons with stars that cover a range of spectral 
types indicate that the relative depths and widths of the 
SiII and HeI features in the V1507 Cyg spectrum are well matched with those 
in the spectrum of the B8 I star $\beta$ Ori if that star were to 
contribute 60\% of the V1507 Cyg light near 6500\AA\ . 
Despite the similarity in spectral type, the brighter 
star in V1507 Cyg is fainter than expected for a 
B supergiant, with M$_V = -2.3$ as opposed to M$_V \sim 
-7.8$ for $\beta$ Ori \citep[]{przetal2006}. The radius of the brighter 
star is also greater than that of a main sequence star with a similar 
effective temperature, but smaller than that of a supergiant. That 
the bright star in V1507 Cyg is underluminous and undersized 
when compared with stars that have evolved in isolation is 
consistent with it having lost mass as a result of interactions 
with its companion. 

	The discrepant nature of the intrinsic luminosity 
and radius of the brighter star would largely disappear if its properties 
are compared with those of a B8 giant, rather than supergiant. However, 
in Section 3 it is shown that a giant luminosity class is not consistent 
with the relative depths of the SiII and HeI lines. The line depths clearly 
indicate that these transitions occur in an environment with a surface 
gravity that is consistent with that of a supergiant.

	Given the low rate of mass exchange and the eccentric orbit, it 
is likely that at the present day the brighter star is not in contact with its 
Roche surface. If it were in contact with its Roche surface then the evolution 
of the system might progress at a rapid pace. To demonstrate 
this consider the models generated by \citet{lajandsil2011}, 
who used a smoothed particle hydrodynamics 
code to examine mass transfer in intermediate mass binaries that have a 
range of orbital eccentricities. While the properties of these models do not 
replicate those of V1507 Cyg (e.g. the stars in the models have masses that 
are a few times lower than those found here for V1507 Cyg), 
there are general results that should have a wide applicability.

	As expected, the rate of mass transfer with time in 
the models is not constant. Mass transfer also does not occur 
as a discrete pulse, but is an extended event, with the peak mass transfer 
rate occuring well after periastron, near phases 0.55 - 0.57. This 
delay in the peak transfer rate is attributed to the travel time needed for 
particles to leave the donor star and assemble around the 
gaining star. Perhaps of greatest importance for V1507 Cyg is that 
the rate of mass transfer in the models grows with each periastron 
passage, as the donor star expands in response to mass lost during 
the preceeding encounter. The simulations follow the systems over only a 
few orbits, and if this trend of increasing mass transfer were to continue 
then it may lead to a rapid depletion of gas from the donor star. 
We do not see evidence of such rapid evolution in the system orbit.

\subsubsection{Envelopes around the stars}

	Features that are attributed to the photosphere of the 
companion star are not detected, as visible light from this object 
appears to be blocked by a thick circumstellar envelope. This envelope 
is the likely source of the mid-infrared emission that originates from 
the system. Still, after subtracting out the spectrum 
from the brighter star there remains an emission spectrum 
that is dominated by SiII and FeII lines, as well as HeI 6678 
in absorption and emission. Based on their wavelength behaviour with phase 
we attribute these features to the envelope around the companion. 
The radial velocities measured from the SiII emission 
and HeI profile are in broad agreement, thereby indicating a 
common kinematic heritage. 

	The envelope around the companion is in a dynamic 
state. The broad HeI trough indicates that the envelope contains 
a component that is rotating and expanding, and there is also 
a red shifted absorption component that is indicative 
of infall. The mean expansion velocity is $216 \pm 5$ 
km/sec if the flow is isotropic, but the actual expansion velocity will be 
greater if there are projection effects that are related to the inclination 
of the system. Variations in the shape of the HeI profile suggest 
that the envelope around the companion is not symmetric, and that the 
distribution of material at a given orbital phase changes with time.

	If there is a dusty medium around 
the companion then light from the companion will 
be re-emitted at longer wavelengths. We have examined the 
colors of V1507 Cyg using photometry from the WISE \citep[]{wrietal2010} 
source catalogue \citep[]{wrietal2019}, and compared these with 
the colors of classic W Serpentis systems, which are thought to 
have thick disks around the gaining stars. 
We find that W1--W2$=0.3 \pm 0.1$ and W1--W4$= 1.4 \pm 
0.1$ for V1507 Cyg. For comparison, the mean colors 
of V367 Cyg, W Ser, SX Cas, RX Cas, and W Cru are $\overline{W1-W2} = 0.4 
\pm 0.1$ and $\overline{W1-W4} = 1.3 \pm 0.3$, where the uncertainties are 
the formal errors in the mean. The MIR spectral energy distribution 
of V1507 Cyg is thus similar to what is seen in other interacting 
binary systems.

	H$\alpha$ emission is a prominent feature in the 
spectrum at red wavelengths. The profile of this emission has a complex 
shape that varies with orbital phase, containing sub-structures that 
suggest at least two components produce the emission. There is also 
a broad, blue-shifted absorption component that is visible at some phases. 

	Evidence has been presented that the bulk of the H$\alpha$ 
emission moves in a manner that tracks the motion of the brighter 
star. There is also evidence that the absorption 
feature moves in a sense that is consistent with the brighter star. 
Orbit-to-orbit variations in the emission are seen after periastron, 
suggesting that the level of activity that powers the emission changes 
with time. Thus, the environment around the stars is not in an 
equilibrium state, but experiences cycle-to-cycle activity.

\subsubsection{Component masses}

	The masses of both components have been estimated by 
assuming that the emission and absorption features that are 
associated with the envelope track the motion of the companion. 
A potential concern is that the radial velocities obtained from 
the SiII emission and HeI P Cygni profiles define a trend with orbital phase 
that departs from that expected near phases 0.3 -- 0.4, although the 
radial velocities at other phases do not appear to be outliers 
(e.g. Figure 7). While the cause of the behaviour of velocities near 
phases 0.3 -- 0.4 is a matter of speculation, one possible cause might be 
distortions in the disk that result from tidal effects post-periastron.

	When compared against the velocities of the brighter star, the 
velocities obtained from SiII emission have less scatter than those measured 
from HeI, and so the SiII velocities were adopted 
for calculating component masses. When combined 
with the amplitude of the velocity curve of the brighter star, the SiII 
emission line velocities yield component masses of $6 \pm 1$ and $14 \pm 1$ 
M$_{\odot}$ with an orbital inclination of 46 degrees, as estimated by 
\citet{ber1998}. A lower limit to the masses can be obtained given that the 
system is not eclipsing. If a maximum inclination of $60^o$ is assumed then 
the limits on the masses are $> 3.8 \pm 0.5$ and $> 8.2 \pm 0.4$ M$_{\odot}$.

	The present study thus finds that 
V1507 Cyg contains stars that, if they were on the main 
sequence, would have late B or earlier spectral types. That the companion 
is more massive than the brighter star is consistent with mass transfer 
from the brighter star -- which was initially the more massive star -- 
progressing to the point where the mass ratio has reversed. 
That the brighter star is less than one half the mass of the 
companion indicates that mass transfer must be well-advanced.

	The total system mass based on this analysis exceeds the 
initial system mass estimated by \citet{dav2023}, who used the photometric 
properties of stars that have distances and proper motions that are 
similar to those of V1507 Cyg. We put forward two 
possible explanations for this result. The first is that the age of the 
moving group associated with V1507 Cyg has been overestimated, presumably 
due to stochastic effects in populating the regions near the 
main sequence turn-off. The second is that V1507 Cyg may not be related to the 
nearby stars that happen to have similar proper motions. V1507 Cyg would 
then have presumably formed in a cluster or association that was disrupted 
early in its evolution \citep[e.g.][]{lad2003}, or has moved 
significantly from its place of origin. If there was a disrupted cluster 
then V1507 Cyg appears to be the lone survivor along this line of sight, 
as the CMD of all objects within 10 parsecs of V1507 Cyg (i.e. with no limits 
based on proper motions) shown in Figure 6b of \citet{dav2023} contains 
no main sequence stars with ages $> 1$ Gyr. As for it moving far from 
its place of origin, the location of V1507 Cyg on the proper motion plane is 
not exceptional \citep[Figure 1b of][]{dav2023}, in the sense that it is 
not an obvious outlier. 

\subsubsection{The system light curve}

	V1507 Cyg does not show the large amplitude 
photometric activity that is seen in the light curves of many CBSs, 
and is most commonly attributed to accretion and the formation of hot spots. 
This might indicate that the mass flow rate is very low at the current epoch, 
and it is not even clear if the brighter star comes into contact with 
its Roche surface during periastron. Even if there 
is significant mass transfer, photometric activity can be curtailed 
if the mass stream has a grazing impact on the accretion disk, with 
the result that a prominent hot spot may not form. 
This has been proposed by \citet{menetal2016} to 
explain the lack of photometric activity in some CBSs.

	Based on the shape and modest amplitude 
of the V1507 Cyg light curve, coupled with the spectral type of the 
brighter star, we suspect that the brighter star may be an $\alpha$ Cyg 
variable, and it is that variability that is responsible for the system 
light curve. $\alpha$ Cyg variables are B or early A supergiants 
that are experiencing non-radial pulsations. The light curves have modest 
amplitudes ($\leq 0.1$ mag) with cycle-to-cycle variations about the mean light 
curve. $\alpha$ Cyg variables also span a range of luminosities. For example, 
the 29 variables identified by \citet[]{waeetal1998} 
have log(L/L${_\odot}$) between 2.78 to over 7. The variables in that sample 
also have periods that range from a few days to over one hundred days. 
If the brighter star in V1507 Cyg is an $\alpha$ Cyg variable then its 
luminosity places it near the fainter end of $\alpha$ Cyg variables.

	The suggestion that the brighter star is an $\alpha$ Cyg variable 
can be checked by searching its spectrum for radial velocity 
variations due to pulsations. A complicating factor is that the 
periodicity of the V1507 Cyg light curve is the same as the system orbit, 
and so tidal effects presumably regulate the pulsation activity. 
Any radial velocity variations due to pulsations would then 
be best detected after removing the radial velocities that are associated 
with the motion of the brighter star in its orbit.

\subsection{The Eccentric Orbit}

	The eccentricity of the orbit is 
a key consideration when discussing the nature 
of V1507 Cyg, as its cause likely holds clues into the 
very recent evolution of the system. Eccentric orbits are not 
expected to be common in CBSs, as mass transfer and tidal effects will 
generally act to circularize the orbits of the components. Empirical evidence 
for the impact of tidal forces on the shape of the orbit is seen in 
relations between eccentricity and the relative sizes 
of stars when measured in the context of the separation of the 
components, as well as with the system age when measured in terms of 
the circularization timescale \citep[e.g. Figures 1 and 8 of][]{may1991}. 
It is because of the eccentric orbit that we have not attempted to 
model the evolution of V1507 Cyg or compare it with existing model 
libraries, since an eccentric orbit suggests that the evolution has been 
influenced by factors other than those usually considered in models of 
CBS evolution. In this section we briefly discuss some of the possible 
causes of an eccentric orbit in an evolved CBS such as V1507 Cyg.

	Models suggest that there are conditions under which mass 
transfer may not result in a rapid circularization of the orbit 
\citep[e.g.][]{sepetal2009}, and that mass transfer may 
even counteract the tendency for tidal forces to 
circularize orbits \citep[e.g.][]{bonetal2008, dosandkal2016}. 
In fact, there are a modest number of interacting 
binary systems that have an eccentric orbit 
\citep[e.g.][]{vosetal2013, bofetal2016}. However, with the exception 
of systems that contain compact objects, these systems 
tend to have orbital periods of many years and large 
eccentricities, and so the components might be subjected to greatly reduced 
tidal effects during a large fraction of their orbital cycles. 
Again with the exception of systems that contain a compact 
object, the author is not aware of an interacting system with stars in an 
eccentric orbit that has an orbital period similar to V1507 Cyg.

	An eccentric orbit might be expected in a 
CBS if it is observed fortuitously during the earliest stages of interaction, 
prior to orbit circularization. However, the properties of V1507 Cyg 
suggest that it is a highly evolved system that has undergone mass transfer 
to the point where the mass ratio has reversed and the gainer is now 
more massive than the donor. Hence, the orbital 
eccentricity is likely not a property that was instilled at birth.

	An eccentric orbit could be the result of an 
interaction with another body. In fact, many 
eclipsing binaries with eccentric orbits have 
a third body in the system \citep[e.g.][]{zasetal2021}, and 
the presence of a third body may play a role in spurring interactions 
between the two stars that are in a tighter orbit \citep[e.g.][]{tooetal2020}.
We note that the astrometric measurements for V1507 Cyg in GAIA DR3 are 
assigned a re-normalized unit weight error (RUWE) 
of 1.5, which means that the parallax is less certain 
than that of isolated stars with the same apparent brightness. 
This could be a consequence of the orbital motions of the stars, but 
might also be indicative of an unresolved third body.

	If a third body were responsible for 
producing an eccentric orbit after mass transfer then it would have to 
be on an orbit that would have allowed mass transfer between the brighter 
star and the companion to proceed uninterrupted. Assuming 
Case B mass transfer then the pace of mass transfer 
is set by the evolutionary timescale of the initially 
more massive star as it evolves off of the main sequence. 
Assuming conservative mass transfer and an initial mass ratio near unity, then 
the mass losing star in V1507 Cyg had an initial mass of $\sim 10$M$_{\odot}$. 
The timescale for the development of a convective envelope in a 10M$_{\odot}$ 
star is $\sim 3 \times 10^4$ years \citep[e.g.][]{vicetal2021}, and the 
timescale for mass transfer prior to the reversal 
of the mass ratio in V1507 Cyg should then be a few times this value. 
Given the evidence that mass transfer has happened in V1507 Cyg then 
the orbital period of any third body should be in excess of this 
timescale, otherwise repeated interactions with the third body would have 
disrupted mass exchange.

	If a third body were present then it would also affect the 
orbital parameters of the binary system, such as inducing changes in the mean 
system velocity $\gamma$, which appears to have stayed constant to within 
$\sim 1$ km/sec over roughly a century. Limits can be placed on the nature of 
a supposed third body that would not change the $\gamma$ 
measurements by more than this amount. The image slicer 
at the entrance to the 1.2 metre spectrograph covers $\sim 2$ 
arcsec, which corresponds to $\sim 2000$ AU at the distance of V1507 Cyg. 
If a third body were within the area sampled by the slicer then it 
would only be detected if it had an intrinsic brightness that was 
within a few magnitudes of the brighter star, and so absorption features 
from stars with masses that are near or below solar would not be detected. 
As no such light is seen in the spectra, then if a third body is 
present within a few arcsec then it must have a mass $\leq 1$M$_{\odot}$. 
Any such third body must then have a characteristic separation of at least 100 
AU from V1507 Cyg in order not to produce a noticeable shift in $\gamma$. 
However, the period of such an object would be roughly two thousand years at a 
distance of 100 AU, and so might disrupt mass transfer. If the 
typical separation is 10$^3$ AU then the period would be 6000 years,
which is again too short. The upshot is that if a third body is 
present then it must have a separation of at least $\sim 10^5$ AU 
from the binary system.

	There is as yet no firm evidence for a companion to the V1507 Cyg 
system that is on a wide orbit. \citet{dav2023} examined the spatial volume 
around V1507 Cyg, and identified a modest sample of stars that have 
proper motions that are similar to V1507 Cyg. These stars have masses of 
no more than 2M$_{\odot}$. However, the closest object to 
V1507 Cyg that has a similar parallax and proper motion is 
$\sim 400$ arcsec away \citep[][]{dav2023}, corresponding to a distance of 
almost $4 \times 10^5$ AU. It is not clear if such a companion would 
remain bound to V1507 Cyg.

	A fluke flyby encounter is another mechanism that 
could cause an eccentric orbit. Based on the distribution of objects 
in proper motion space, \citet{dav2023} found a loose stellar cluster 
within $\sim 15$ pc of V1507 Cyg. The projected distribution 
of stars in the volume of space around V1507 Cyg suggests that 
it is located near the edge of this cluster, with a proper motion 
that differs from that of the main body of the cluster. If V1507 
Cyg had an encounter with a cluster member then such an interaction 
may have induced an eccentric orbit.

	The final possibility that we consider is that the eccentric 
orbit may have been triggered by a process internal to the system, 
such as a supernova explosion. As discussed in Section 7, there is evidence for 
structure in the ISM that is centered on V1507 Cyg. If this structure 
is a SNR associated with V1507 Cyg then it must be 
relatively young. The HI bubble has a radius of $\sim 20$ 
arcmin, and so would have a physical radius of $\sim 6$ parsecs if at the 
distance of V1507 Cyg. Assuming that SNR expansion velocities fall in the range 
$10^3 - 10^4$ km/sec, then the SNR would have an age between 600 ($10^4$ 
km/sec) to 6000 ($10^3$ km/sec) years. SNR may persist for up to $\sim 0.4$ Myr 
\citep{suzetal2021}, and the age of the proposed SNR 
fits well within the expected lifetime. Moreover, as this would have been a 
very recent event then the system orbit may not have had 
time to circularize. The proper motion of V1507 Cyg also suggests that 
over a 6000 year timespan the system would have moved $\sim 
0.5$ arcmin on the sky, and so would still appear more-or-less centered on the 
SNR in Figure 11. Still, we emphasize that the association of the HI structure 
with V1507 Cyg is not ironclad, as the line of sight passes 
through a densely populated part of the Galactic Disk, making 
depth effects a major concern. 

	If there is a SNR associated with V1507 Cyg then what was the 
progenitor? One possibility is that V1507 Cyg was part of a hierarchical 
system, and the progenitor was a third body that was well separated from the 
present day V1507 Cyg system and was initially more massive than either of the 
components in the binary system. Hierarchical triple systems are common 
among massive stars \citep[e.g.][]{sanetal2014}, and dynamical measurements 
have hinted at the presence of a black hole around some binary systems 
	\citep[e.g.][]{qiaetal2008, lia2010, menetal2021}, 
that presumably originated from a now defunct massive 
star. If there was a such a massive companion to the V1507 Cyg system 
then the stability of the ephemeris suggests it must be at a 
distance where it does not perturb the orbit of the V1507 Cyg system at the 
present day. Perhaps it was ejected entirely from the system by a SN kick 
that could result from an asymmentric explosion.

	Could the companion to the brighter star 
in the V1507 Cyg system be a compact object, and hence the progenitor 
of the SNR? The mass estimated here for the companion is 
close to, but still within, the upper limit of the 
mass range expected for stellar black holes formed from 
a solar metallicity progenitor \citep[]{beletal2010}. Nevertheless, 
if the companion is a collapsed object then its progenitor would have 
initially been the more massive star in the system. We might then expect to 
see evidence of processed material in the spectrum of that star, and this 
should be the subject of future work. In any event, 
given that the mass of the black hole is near the upper limit of that observed 
among solar metallicity objects then the progenitor would have likely been 
one of the most massive stars in the Galaxy, making it an exceptional 
object for this part of the sky, as star-forming activity is in the 
foreground well behind V1507 Cyg. V1507 Cyg is also not a source of the x-ray 
emission, at least at the present day, that is associated with 
many black hole candidates. Still, while x-ray emission is 
expected as material is accreted onto a black hole, 
an opaque disk might block it from the line of sight. 
As the viewing angle of the system is not edge-on then any material that would 
block emission from the companion must engulf a large fraction of that star 
-- even near the poles -- rather than being restricted to a disk 
in the orbital plane. The P Cygni-like profile 
of HeI 6678 suggests that such an envelope 
might be present. Accretion onto a compact object might also 
be hindered if the disk is rotating rapidly, as has been suggested 
for the MWC 656 system by \citet{casetal2014}.

\subsection{Wrapping Up}

	Evidence has been presented that V1507 Cyg is an evolved 
binary system where there has been significant mass transfer. 
The findings that support this claim are that (1) the 
brighter star at visible wavelengths is 
underluminous and undermassive when compared with stars 
that have a similar spectral type, and (2) the brighter 
star is less than one half the mass of the obscured 
companion. Mass transfer has thus progressed for an 
extended period of time, and the mass ratio of the system has changed 
substantially. Assuming an initial mass ratio 
close to unity and conservative mass transfer, then 
the total mass of the system at the present day suggests that the 
components initially had a mid-B spectral type. 
The companion at the present day is likely a 
heavily obscured, moderately massive star. 

	The mass flow has presumably subsided at the present epoch, given 
that the eccentric orbit limits mass flow to times near periastron. 
In fact, it is not clear if the brighter star comes into contact with 
its Roche surface during periastron, and so the reservoir available for 
mass transfer may be limited to an extended envelope around that star. 
Still, the components are tidally interacting, as the signatures 
of periastron effects are seen in the spectra of the 
brighter star and the envelope around the companion.

	A tacit assumption in the discussion has been that the orbits of the 
stars during much of the system's past were circular. 
However, explaining the present day orbital eccentricity 
is problematic, as none of the mechanisms considered here to explain 
the eccentric orbit are problem-free. It is the opinion of the 
author that V1507 Cyg was initially part of a hierarchical system. The stars in 
the V1507 Cyg system were in a circular orbit when the 
more massive third body went SN. It was the recent demise 
of that third object, and its possible ejection from the system, 
that perturbed the orbit of the V1507 Cyg system. 
The emission seen at 408 MHz and 1420 MHz that is centered on the system 
is the presumed remnant of that event. 

	There are numerous avenues for future work. 
As a CBS with an eccentric orbit, V1507 Cyg is an important 
system for examining periastron effects. A comparison with model line 
profiles such as those constructed by \citet{moretal2005} but for 
system parameters that are more appropriate for V1507 Cyg would be useful to 
determine the impact of tidal effects on the brighter star as well as 
the envelope around the companion. 

	Monitoring the behaviour of the system spectrum to check 
for periodic and non-periodic behaviour in line characteristics, especially 
before and after periastron, will be rewarding. Such observations would 
provide additional information that is useful for monitoring the 
mass flow at a point in the orbit when it is expected to peak. 
Observations near phases 0.3 and 0.4 are of particular interest to 
assess the behaviour of radial velocities and mass flow with time at 
that point in the orbit.

	Spectra that cover a wider wavelength range could also be used to 
search for abundance anomalies in the chemical content 
of the brighter star. A non-solar chemical mixture might be expected if 
core processed material has been brought to the surface due to mixing 
within the star, which might be driven by tidal effects at the present day, 
or could be present if another star transfered mass to the B8 I star 
at some point in the past. Spectra of this nature could also be used 
to assess the proposal that the B8 I star is an $\alpha$ Cyg variable 
by searching for radial velocity variations associated with pulsations. If 
evidence for pulsations is found then it might be possible to 
probe the interior structure of the brighter star.

	Further examination of the material around the (as yet undetected) 
companion over a broad wavelength range is also of interest to 
provide additional insights into the system mass ratio, the 
nature of the envelope, and the rate of mass accretion/ejection. The latter 
is of interest as the HeI 6678 profile in the residual 
spectrum suggests that there is mass outflow from the environment 
around the companion, and so only a fraction of the mass transfered from 
the B8 I star may eventually be deposited onto the companion at the 
present day. Monitoring the system at x-ray wavelengths 
might also provide insights into the rates of mass deposition onto the 
companion and place limits on its strength. Given the eccentric orbit then 
such emission might vary with orbital phase.

	Finally, studies of the environment around the system 
will also provide clues into its past. For example, is the HI 
structure detected in the CGPS images a SNR? If so, is it at the same 
distance as V1507 Cyg? Answers to these questions have the potential 
to provide insights into the origins of the eccentric orbit.

\acknowledgements{It is a pleasure to thank David Bohlender for 
providing an initial reduction of the spectra used 
in this paper. Thanks are also extended to Dmitry Monin for his tireless 
efforts working with the DAO telescopes. Sincere thanks are also extended 
to the anonymous referee for providing suggestions that greatly improved the 
presentation and interpretaton of the results. This paper is based mainly on 
observations obtained at the Dominion Astrophysical 
Observatory, NRC Herzberg, Programs in Astronomy 
and Astrophysics, National Research Council of Canada. This research 
also has made use of the NASA/IPAC Infrared Science
Archive (https://doi.org/10.26131/irsa142), 
which is funded by the National Aeronautics and Space Administration and
operated by the California Institute of Technology.
This work has also made use of data from the European Space Agency (ESA) mission
{\it Gaia} (\url{https://www.cosmos.esa.int/gaia}), processed by the {\it Gaia}
Data Processing and Analysis Consortium (DPAC,
\url{https://www.cosmos.esa.int/web/gaia/dpac/consortium}). Funding for the DPAC
has been provided by national institutions, in particular the institutions
participating in the {\it Gaia} Multilateral Agreement. Finally, the research 
presented in this paper has also used data from the Canadian Galactic 
Plane Survey, a Canadian project with international partners, supported 
by the Natural Sciences and Engineering Research Council.}


\parindent=0.0cm
\clearpage

\begin{thebibliography}{}

\bibitem[Baines et al. (2018)]{baietal2017}
Baines, E. K., Armstrong, J. T., Schmidt, H. R. et al. 2018, \aj, 155, 30

\bibitem[Belczynski et al. (2010)]{beletal2010}
Belczynski, K., Bulik, T., Fryer, C. L., Ruiter, A., Valsecchi, F., Vink, J. S., \& Hurley, J. R. 2010, \apj, 714, 1217

\bibitem[Berdyugin \& Tarasov (1998)]{ber1998}
Berdyugin, A. V., \& Tarasov, A. E. 1998, AstL, 24, 111

\bibitem[Boffin et al. (2016)]{bofetal2016}
Boffin, H. M. J., Hillen, M., Berger, J. P., et al. 2016, \aap, 564, A1

\bibitem[Bonacic Marinovic et al. (2008)]{bonetal2008}
Bonacic Marinovic, A. A., Glebbeek, E., \& Pols, O. R. 2008, \aap, 480, 797

\bibitem[Broz et al. (2021)]{broetal2021}
Broz, M., Mourard, D., Budaj, J., et al. 2021, \aap, 645, A51

\bibitem[Burdge et al. (2019)]{buretal2019}
Burdge, K. B., Couglin, M. W., Fuller, J. et al. 2019, \nat, 571, 528

\bibitem[Burdge et al. (2020)]{buretal2020}
Burdge, K. B., Couglin, M. W., Fuller, J. et al. 2020, \apjl, 905, L7

\bibitem[Casares et al. (2014)]{casetal2014}
Casares, J., Negueruela, I., Ribo, M., Ribas, I., Paredes, J. M., Herrero, A., \& Simon-Diaz, S. 2014, \nat, 505, 378

\bibitem[Clark et al. (2013)]{claetal2013}
Clark, J. S., Bartlett, E. S., Coe, M. J. et al. 2013, \aap, 560, A10

\bibitem[Davidge (2023)]{dav2023}
Davidge, T. J. 2023, \aj, 165, 189

\bibitem[Deschamps et al. (2015)]{desetal2015}
Deschamps, R., Braun, K., Jurissen, A. et al., 2015, \aap, 577, A55
 
\bibitem[de Vaucouleurs (1978)]{dev1978}
de Vaucouleurs, G. 1978, \apj, 233, 351

\bibitem[Dosopoulou \& Kalogera (2016)]{dosandkal2016}
Dosopoulou, F., \& Kalogera, V. 2016, \apj, 825, 70

\bibitem[El-Badry \& Qataert (2021)]{elb2021}
El-Badry, K. \& Qataert, E. 2021, \mnras, 502, 3436

\bibitem[GAIA Collaboration (2022)]{gai2022}
Gaia Collaboration, Vallenari, A., Brown, A. G. A., et al. 2022, arXiv:2208.00211

\bibitem[Hamman \& Persson (1992)]{ham1992}
Hamann, F. \& Persson, S. E. 1992, \apjs, 82, 285

\bibitem[Hill et al. (1976)]{hil1976}
Hill, G., Hilditch, R. W., \& Pfannenschmidt, E. L. 1976, PDAO, 15, 1

\bibitem[Hutchings \& Laskarides (1972)]{hut1972}
Hutchings, J. B., \& Laskarides, P. G. 1972, \mnras, 155, 357

\bibitem[Hutchings \& Redman (1973)]{hut1973}
Hutchings, J. B., \& Redman, R. O. 1973, \mnras, 163, 209

\bibitem[Kobayashi et al. (2020)]{kobetal2020}
Kobayashi, C., Karakas, A. J., \& Lugaro, M. 2020, \apj, 900, 179

\bibitem[Lada \& Lada (2003)]{lad2003}
Lada, C. J., \& Lada, E. A. 2003, \araa, 41, 57

\bibitem[Lajoie \& Sills (2011)]{lajandsil2011}
Lajoie, C-P \& Sills, A. 2011, \apj, 726, 67

\bibitem[Liao \& Qian (2010)]{lia2010}
Liao, W.-P. \& Qian, S.-B. 2010, \mnras, 405, 1930

\bibitem[Martin et al. (2009)]{maretal2009}
Martin, R. G., Tout, C. A., \& Pringle, J. E. 2009, \mnras, 397, 1563

\bibitem[Mayer \& Hanna (1991)]{may1991}
Mayer, P., \& Hanna, M A.-M. 1991, BAICz, 42, 98

\bibitem[Meng et al. (2021)]{menetal2021}
Meng, G., Zhang, L.-Y., Han, X. L., et al. 2021, \mnras, 503, 324

\bibitem[Mennikent et al. (2016)]{menetal2016}
Mennikent, R. E., Otero, S., \& Kolaczkowski, Z. 2016, \mnras, 455, 1728

\bibitem[Merrill (1949)]{mer1949}
Merrill, P. W. 1949, \apj, 110, 59

\bibitem[Monin et al. (2014)]{monetal2014}
Monin, D., Saddlemyer, L., \& Bohlender, D. 2014, \rmxaa, 45, 69

\bibitem[Moreno et al. (2005)]{moretal2005}
Moreno. E., Koenigsberger, G., \& Toledano, O. 2005, \aap, 437, 641

\bibitem[Moreno et al. (2010)]{moretal2010}
Moreno. E., Koenigsberger, G., \& Harrington, D. M.. 2010, \aap, 528, A48

\bibitem[Morgan \& Keenan (1973)]{mor1973}
Morgan, W. W., \& Keenan, P. C. 1973, \araa, 11, 29

\bibitem[Morton (1960)]{mor1960}
Morton, D. C. 1960, \apj, 132, 146

\bibitem[Paunzen et al. (2006)]{pauetal2006}
Paunzen, E., Netopil, M., Iliiev, I. Kh., Maitzen, H. M., Claret, A., \& Pintado, O. I. 2006, \aap, 454, 171
 
\bibitem[Pavlovski et al. (1979)]{pavetal1979}
Pavlovski, K., Harmanec, P., Horn, J., Koubsky, P., Zdarsky, F., \& Kriz, S. 1979, IBVS, 1689, 1

\bibitem[Plavec \& Koch (1978)]{pla1978}
Plavec, M., \& Koch, R. H. 1978, IBVS, 1482, 1

\bibitem[Przybilla et al. (2006)]{przetal2006}
Przybilla, N., Butler, K., Becker, S. R., \& Kudritzki, R. P. 2006, \aap, 445, 1099

\bibitem[Qian et al. (2008)]{qiaetal2008}
Qian, S.-B., Liao, W.-P., \& Fernandez Lajus, E. 2008, \apj, 687, 466

\bibitem[Ren et al. (2021)]{renetal2021}
Ren, F., Chen, X., Zhang, H., et al. 2021, \apjl, 911, L20

\bibitem[Rucinski (1997)]{ruc1997}
Rucinski, S. M. 1997, \aj, 113, 407

\bibitem[Sana et al. (2014)]{sanetal2014}
Sana, H., Le Bouquin, J.-B., Lacour, S., et al. 2014, \apjs, 215, 15

\bibitem[Sanchez-Blazquez et al. (2006)]{sanetal2006}
Sanchez-Blazquez, P., Peletier, R. F., Jimenez-Vicente, J. et al. 2006, \mnras, 371, 703

\bibitem[Sepinsky et al. (2009)]{sepetal2009}
Sepinsky, J. F., Willems, B., Kalogera, V., \& Rasio, F. A. 2009, \apj, 702, 1387

\bibitem[Smak (1962)]{sma1962}
Smak, J. 1962, \actaa, 12, 28

\bibitem[Smith \& Tombleson (2015)]{smi2015}
Smith, N., \& Tombleson, R. 2015, \mnras, 447, 598

\bibitem[Stassun \& Torres (2021)]{sta2021}
Stassun, K. G., \& Torres, G. 2021, \apjl, 907, L33

\bibitem[Stepien (2006)]{ste2006}
Stepien, K. 2006, \actaa, 56, 199

\bibitem[Suzuki et al. (2021)]{suzetal2021}
Suzuki, H., Bamba, A., \& Shibata, S. 2021, \apj, 914, 103

\bibitem[Taylor et al. (2003)]{tayetal2003}
Taylor, A. R., Gibson, S. J., Peracaula, M. et al. 2003, \aj, 125, 3145

\bibitem[Toonen et al. (2020)]{tooetal2020}
Toonen, S., Portegies Zwart, S., Hamers, A. S., \& Bandopadhyay, D. 2020, \aap, 640, A16

\bibitem[Torres et al. (2010)]{toretal2010}
Torres, G., Andersen, J., \& Gimenez, A. 2010, \aapr, 18, 67

\bibitem[Trimble \& Thorne (1969)]{tri1969}
Trimble, V. \& Thorne, K. 1969, \apj, 156, 1013

\bibitem[Vick et al. (2021)]{vicetal2021}
Vick, M., MacLeod, M., Lai, D., \& Loeb, A. 2021, \mnras, 503, 5569

\bibitem[Vos et al. (2013)]{vosetal2013}
Vos, J., Ostensen, R. H., Nemeth, P., et al. \aap, 559, A54

\bibitem[Waelkens et al. (1998)]{waeetal1998}
Waelkens, C., Aerts, C., Kestens, E., Grenon, M., \& Eyer, L. 1008, \aap, 330, 215

\bibitem[Wilson et al. (1984)]{wiletal1984}
Wilson, R. E., Rafert, J., \& Markworth, N. L. 1984, IAPPP Comm., 16, 1

\bibitem[Wright et al. (2010)]{wrietal2010}
Wright, E. L., Eisenhart, P. R. M., Mainzer, A. K., et al. 2010, \aj, 140, 1868

\bibitem[Wright et al. (2019)]{wrietal2019}
Wright, E. L., Eisenhart, P. R. M., Mainzer, A. K., et al. 2019, WISE All-Sky Source Catalog (Pasadena, CA: IPAC)

\bibitem[Zahn (1977)]{zah1977}
Zahn, J.-P. 1977, \aap, 57, 383

\bibitem[Zasche et al. (2021)]{zasetal2021}
Zasche, P., Henzl, Z., \& Masek, M. 2021, \aap, 652, A81

\end{thebibliography}
\end{document}